\documentclass[12pt]{article}
\usepackage[english]{babel}
\usepackage{amsfonts,amsmath,mathrsfs,stmaryrd,amsthm}
\usepackage{graphicx}
\usepackage{amssymb}
\usepackage{rotating}
\usepackage{amsmath}
\newcommand{\tr}{^{\prime}}
\usepackage{verbatim} 
\usepackage{tabularx}
\usepackage{natbib}
\usepackage{caption}
\usepackage{color}
\usepackage{hyperref}
\usepackage{url}
\usepackage{mathtools}
\usepackage{booktabs}
\usepackage{multirow}
\usepackage{epstopdf}
\usepackage{epsfig}
\usepackage{soul,color}
\usepackage{bm}
\usepackage{latexsym, graphicx, alltt}
\usepackage[table]{xcolor}
\usepackage{float}
\usepackage[caption=false]{subfig}

\renewcommand{\hat}{\widehat}
\renewcommand{\tilde}{\widetilde}
\renewcommand{\bar}{\overline}

\def\b#1{\mbox{\boldmath $#1$}}
\newcommand{\bu}{\bar{u}}
\def\bl#1{\mbox{\scriptsize \boldmath {$#1$}}} % - \bl: grassetto per apice
\newcommand{\be}{\beta}

\renewcommand{\th}{\theta}

\def\E{{\rm E}}    % - \bt: grassetto in formula e poi trasposto
\def\bSig\mathbf{\Sigma}

\newcommand{\Var}{\mathrm{Var}}

\def\baselinestretch{1.5}\def\arraystretch{0.75}
\title{Maximum likelihood estimation of hidden Markov models
for continuous longitudinal data with missing responses and dropout}
\author{Silvia Pandofi \\
{\small University of Perugia (IT)} \\
{\small silvia.pandolfi@unipg.it}
\and
Francesco Bartolucci \\
{\small University of Perugia (IT)} \\
{\small francesco.bartolucci@unipg.it}
\and
Fulvia Pennoni \\
{\small University of Milano-Bicocca (IT)} \\
{\small fulvia.pennoni@unimib.it}
}
\begin{document}

\label{firstpage}

\maketitle
\vspace*{-0.5cm}
\def\baselinestretch{1.1}
\vspace*{-0.5cm}
\def\baselinestretch{1.1}

\begin{abstract}
We propose an inferential approach for maximum likelihood estimation of the hidden Markov models for continuous responses. We extend  to the case of longitudinal observations the finite mixture model of multivariate Gaussian distributions with Missing At Random (MAR) 
outcomes, also accounting for possible dropout.
The resulting hidden Markov model accounts for different types of missing pattern: ($i$) partially missing outcomes at a given time occasion; ($ii$) completely missing outcomes at a given time occasion (intermittent pattern); ($iii$) dropout before 
the end of the period of observation (monotone pattern). 
The MAR assumption is formulated to deal with the first two types of missingness, while to account for informative dropout 
we assume an extra absorbing state. 
Maximum likelihood estimation of the model parameters is based on an extended %SP: extended
Expectation-Maximization algorithm relying on suitable recursions.
The proposal is illustrated by a Monte Carlo simulation study and an application based on  historical data on primary biliary cholangitis.  
\vskip5mm
\noindent {\sc Keywords:} Expectation-Maximization algorithm; Forward-backward recursion; Latent Markov model; Missing values; Prediction

\end{abstract}
\section{Introduction}\label{intro}
It is well known that longitudinal data \citep{mole:verb:00,digg:02,fitz:garr:lair:ware:04,hsia:05} %SP: spostata prima citazione del 2004
 are frequently affected by missing data that may arise in different ways  \citep{little:rub:20}.
The missing pattern is defined as monotone when individuals may drop out 
from the sample before the end of the study.
In medical studies, this may be  due to a terminal event, such as the death of a patient.
Another type of missingness, also referred to as intermittent, is when individuals are still in the sample, but for any reason,  they do not provide responses at one or more time  occasions. 
This is the case when a clinical visit is missed.
The problem is particularly severe when the missing data mechanism is informative  \citep{litt:95,albe:foll:wang:suh:02} or not ignorable \citep{rub:76,albe:00,little:rub:20}, that is, when the missingness depends on unobserved variables \citep{diggle:ken:94}. 

The literature on statistical methods for handling missing data with repeated measurements  is relatively wide. In this regards, as recently summarized in \cite{zhou:20}, we can distinguish between different  approaches, which include selection models   \citep{mole:etal:97,maru:15}, pattern-mixture models \citep{litt:94, follmann:95,kenw:etal:03, mar:alf:20}, and shared-parameter (or joint) models \citep{wulf:tsia:97,  hend:etal:00, hsie:06,  rizo:12,bartolucci2015discrete,lange:et:al:15, zhang:20}. 

In this paper, we  extend to the case of longitudinal observations the Finite Mixture Model (FMM) of Gaussian distributions \citep{titt:smith:mark:85,mcla:peel:00} under the Missing At Random (MAR) assumption \citep{Hunt:Jorgensen:2003, dizi:07,eiro:14, dela:18} by proposing an inferential approach to obtain exact maximum likelihood estimates of model parameters. 
A Hidden Markov Model \citep[HMM;][]{bart:farc:penn:13,Zucchini2016} for multivariate continuous outcomes results, which is 
based on  conditional Gaussian distributions and 
can address both monotone and intermittent missing data patterns. 
The hidden Markov approach is of particular interest when dealing with longitudinal data \citep{bart:farc:penn:14} as it allows us to model time dependence in a flexible way and to perform a dynamic model-based clustering \citep{bou:et:al:19}.
The same individual may move between clusters across time, and these dynamics are provided in terms of trajectories. 
This is because a sequence of discrete latent variables, rather than a single latent variable, is associated with every individual, giving rise to a hidden process assumed to follow a Markov chain, generally of first-order.
The states of this chain correspond to latent clusters or subpopulations of homogenous individuals.

Overall  we propose an extended HMM that takes explicitly into account the following patterns of missing data: ($i$)  partially missing outcomes at a given time %SP: time 
occasion; ($ii$) completely missing outcomes at one occasion without dropout from the sample of the individual; ($iii$)  dropout from the sample. 
The first two cases correspond to the intermittent missing responses defined above, whereas the third corresponds to a monotone missing pattern also defined  as attrition.
In particular, cases ($i$) and ($ii$) are dealt with under the MAR  assumption, according to which the missing pattern  is independent of the missing responses given the observed data. 
In case ($iii$) dropout is not ignorable, 
and specifying a model for the missing data mechanism is in order.
Along with the proposal of  \cite{Montanari2018} and \cite{spag:et:al:11}, %SP: , 
we include a hidden (or latent)  state, which is absorbing, in addition to the other states representing unobserved heterogeneous populations 
that may arise in the study. 

To estimate the proposed  model we rely on the maximum likelihood approach, paying particular attention to the computational aspects.
In more detail, we propose an extended 
Expectation-Maximization \citep[EM;][]{baum:et:al:70,  demp:lair:rubi:77, welch:2003} algorithm based on suitable  recursions.
The estimation algorithm is also employed when there are available covariates supposed to affect the distribution of the latent process  and, in particular,  
the initial and the transition probabilities of the Markov chain  \citep{bart:farc:penn:14}. 
In this way, it is possible to understand the influence of the covariates on the dynamic allocation of the individuals between states over time. 

In order to illustrate the proposed approach, we rely on a series of simulations that also allow us to assess the maximum likelihood estimator in terms of finite sample properties. 
We also show an application based on historical data about primary biliary cholangitis (or cirrhosis) collected by the Mayo Clinic from January 1974 to May 1984 \citep{murt94}.
Data are referred to  several biochemical measurements of the liver function scheduled for the patients according to the  clinical protocol over six months or one year after the first visit. 
Continuous and binary  covariates related to the patients are also available such as age, gender, and medication use. 
The data are very sparse due to missing visits and  the fact that some biochemical measurements were not collected  at each visit. 
Moreover,  dropout occurred due to death. 
The code implemented to estimate the proposed model in  \texttt{R} \citep{R:21}
is based on the package {\tt LMest} \citep{bart:pand:penn:17}, and  it is available on the GITHUB page at the following link ({\em it will follow}).

The remainder of the paper is organized as follows.
In Section \ref{sec:prel}  we recall the FMMs of Gaussian distributions with missing data under the MAR assumption.
In Section \ref{sec:HMmis} we show the proposed HMM formulation. 
In Section \ref{sec:inf}, we outline the inferential approach proposed for estimating model parameters and aspects related to the computation of the standard errors, model selection, and decoding.
In Section \ref{sec:sim} we present the simulation study.
In  Section \ref{sec:appl}, we  illustrate the applicative example, whereas  in Section \ref{sec:concl} we provide some conclusions. 
A supplementary file provides additional information on the data and results.
\section{Preliminaries}\label{sec:prel}
In this section we outline the FMM of Gaussian distributions under 
the MAR assumption and with possible 
covariates. 
\subsection{Model formulation}
We consider individual vectors $\b Y_i = (Y_1,\ldots,Y_r)\tr$ of $r$ continuous response variables, with $i=1,\ldots,n$. 
It is well known that the 
FMM of Gaussian distributions assumes that there exist $k$ components identified by the discrete latent variable $U_i$ such that
\[
\b Y_i|U_i=u\sim N(\b\mu_u,\b\Sigma_u),\quad \quad u=1,\ldots,k.
\]
In the previous expression, $\b\mu_u$ and $\b\Sigma_u$ denote the mean vector and the variance-covariance matrix for the same mixture component $u$, respectively. 
Note that the variance-covariance matrices may be assumed to be constant across the components under the assumption 
of homoschedasticity, which also avoids certain estimation problems \citep[][Chapter 3.8]{mcla:peel:00}. 
More articulated constraints on these matrices may be expressed, as proposed in \cite{Banfield:Raftery:1993}, on the basis of suitable matrix decompositions. 

With missing data it is convenient to partition each response vector as $(\b Y_i^o, \b Y_i^m)\tr$, where $\b Y_i^o$ is the vector of observed variables and $\b Y_i^m$ is that of missing variables.
Using a straightforward notation, the conditional mean vectors and variance-covariance matrix may be decomposed as 
\begin{equation} 
\b\mu_u =  \begin{pmatrix}
\b \mu_u^{o} \\
\b \mu_{u}^m  
\end{pmatrix}, \quad \b\Sigma_u =  \begin{pmatrix}
\b\Sigma_u^{oo} &	\b\Sigma_u^{om}\\
\b\Sigma_u^{mo} &\b\Sigma_u^{mm} \\
\end{pmatrix}, \label{Eq:dec_mu_Si}
\end{equation}
where, for instance, $\b\Sigma_u^{om}$ is the block of $\b\Sigma_u$ containing the covariances between each observed and missing response.
In this way, for the observed responses we have that
\[
\b Y_i^o|U_i=u\sim N(\b\mu_u^o,\b\Sigma_u^{oo}),\quad u=1,\ldots,k.
\]

Without individual covariates, each latent variable $U_i$ has the same distribution based on the component weights $\pi_u=p(U_i=u)$, $u=1,\ldots,k$. 
Consequently, the manifest distribution of the observed responses is given by 
\begin{equation}
f(\b y_i^o) = \sum_{u=1}^k \pi_u\phi(\b y_i^o; \b\mu_u^o, \b \Sigma_u^{oo}),\label{eq:fin_mix}
\end{equation}
where $\b y_i^o$ denotes a realization of $\b Y_i^o$  and $\phi(\cdot; \cdot)$ 
denotes the multivariate Gaussian probability density function.

Individual covariates may be included in the model in different ways.
In particular, we consider the
case of  
component weights affected by these covariates. 
Let $\pi_{iu}=p(U_{i}=u|\b x_i)$  denote the class weight for component $u$ which is now specific of individual $i$ and 
$\b x_i$ the vector of individual covariates, with $i=1,\ldots,n$.
We assume a multinomial logit model of type 
\begin{equation}
\log\frac{\pi_{iu}}{\pi_{i1}}=\b x_i'\b\be_u,\quad u=2,\ldots,k, 
\label{eq:cov}
\end{equation}
where $\b\be_u$ is a vector of regression parameters.
Expression \eqref{eq:fin_mix} for the manifest distribution is obviously extended as
\[
f(\b y_i^o) = \sum_{u=1}^k \pi_{iu}\phi(\b y_i^o; \b\mu_u^o, \b \Sigma_u^{oo}).
\]
\subsection{Expectation-Maximization algorithm with missing data}\label{sec:EM_fin_mix}
In the presence of missing data, the observed log-likelihood can be written as
\[
\ell(\b\theta) = \sum_{i=1}^n \log f(\b y_i^o) = \sum_{i=1}^n \log \left[
\sum_{u=1}^k\pi_u \phi(\b y_i^o; \b\mu_u^o, \b \Sigma_u^{oo})\right].
 \]
Its maximization  for parameter estimation  relies on the EM algorithm \citep{demp:lair:rubi:77}. 
This algorithm is based on the {\em complete-data log-likelihood} that has expression
\[
\ell^*(\b\theta) = \sum_{i=1}^n\sum_{u=1}^k z_{iu} \log [\pi_u  \; \phi(\b y_i; \b\mu_u, \b \Sigma_u)],
\]
where $z_{iu}$ is a binary 
variable indicating whether or not unit $i$ comes from component $u$.

As usual, the EM algorithm alternates two steps until convergence. 
The {\em E-step} consists in computing the conditional expectation of the indicator 
variables given the observed data and the current value of the parameters, that is,
\begin{equation}
\hat{z}_{iu} = \frac{\pi_u \phi(\b y_i^o; \b\mu_u^o, \b \Sigma_u^{oo})}
{\sum_{v=1}^k \pi_v
\phi(\b y_i^o; \b\mu_v^o, \b \Sigma_v^{oo})}.\label{eq:ppost} %SP: messa somma su v visto che u la usiamo a numeratore
\end{equation}
With individual covariates, this rule is modified by substituting every $\pi_u$ with $\pi_{iu}$.

Following the proposals in \cite{eiro:14} and \cite{dela:18}, in order to account for missing data this step also includes the computation of the following conditional expectations
\begin{equation}
\E(\b Y_{i} | \b y_{i}^o, u) =\left(\begin{matrix}
\b  y_{i}^o \\
\E(\b Y_{i}^m | \b y_{i}^o, u)
\end{matrix}\right),
\label{eq:cond_mean} 
\end{equation}
where
\[
\E(\b Y_{i}^m | \b y_{i}^o, u) = \b \mu_{u}^m + \b\Sigma_u^{mo} (\b\Sigma_u^{oo})^{-1} (\b y_{i}^o- \b \mu_{u}^o),
\]
and of the conditional variances
\begin{eqnarray*}
\begin{split}
\Var(\b Y_{i}| \b y_{i}^o, u)  = \left(\begin{matrix}
\b O & \b O \\
\b O & \b \Sigma_u^{mm} - \b \Sigma_u^{mo}(\b\Sigma_u^{oo})^{-1}   \b \Sigma_u^{om}
\end{matrix}\right),
\end{split}
\end{eqnarray*}
where $\b O$ is a matrix of zeros of suitable dimension.
The above expressions originate from the fact that the missing values' conditional distribution also follows a multivariate Gaussian  distribution \citep{ander:03}. 

The {\em M-step} consists in updating the model parameters by maximizing the expected value of  $\ell^*(\b\theta)$ obtained at the E-step.
This maximization leads to the following updating rules for the mean vector and the variance-covariance matrices:
\begin{eqnarray}\label{eq:mu}
\b \mu_u &= &\frac{1}{\sum_{i=1}^n \hat{z}_{iu}}\sum_{i=1}^n \hat{z}_{iu} \E(\b Y_{i} | \b y_{i}^o, u),\\ \label{eq:Si}
\b \Sigma_u  &=& \frac{1
}{\sum_{i=1}^n \hat{z}_{iu}}\sum_{i=1}^n \hat{z}_{iu}\bigg\{ 
\Var(\b Y_{i}| \b y_{it}^o,u) + [\E(\b Y_{i}| \b y_{i}^o, u) -\b \mu_u]  [\E(\b Y_{i}| \b y_{i}^o, u) -\b \mu_u ]\tr  \bigg\} .
\end{eqnarray}
The rule in \eqref{eq:Si} is suitably modified by summing the numerator for all $u$, and dividing the sum 
by $n$, under the constraint of homoschedasticity.
The component weights are updated as
\[
\pi_u = \frac{\sum_{i=1}^n \hat{z}_{iu}}{n},  \quad \quad \quad u =1,\ldots,k,\label{eq:update_piu}
\]
whereas with individual covariates we maximize the complete log-likelihood component
\[
\sum_{i=1}^n\sum_{u=1}^k\hat{z}_{iu}\log\pi_{iu}
\]
with respect to the parameter vectors 
$\b\be_2,\ldots,\b\be_{k}$ defined   in Expression (\ref{eq:cov}) by a Newton-Raphson algorithm. 

These two steps are repeated until convergence, which is checked on the basis of the relative log-likelihood difference, that is,
\[
\frac{\mid\ell(\b\theta^{(s)}) - \ell(\b\theta^{(s-1)})\mid}{\mid \ell(\b\theta^{(s)})\mid} \leq \epsilon,
\]
where $\b\theta^{(s)}$ is the vector of  parameter estimates %SP: s
 obtained at the end of the $s$th M-step 
and $\epsilon$ is a suitable tolerance level (e.g., 10$^{-8}$). A crucial aspect is that of initialization of the algorithm to deal with the likelihood multimodality.
In this regard, it is important to rely on different  
strategies, even based on random rules for choosing the model parameters' starting values. 

\subsection{Model selection and clustering}
Selection  of the number of components   
is an important aspect in applying FMMs 
\citep[][Chapter 6]{mcla:peel:00}.
Typically, this choice is based on information criteria such as the Akaike Information Criterion \citep[AIC;][]{aka:73} or the 
Bayesian Information Criterion  \citep[BIC;][]{sch:78}, obtained through penalizations of the maximum log-likelihood.
In particular, the AIC index is expressed as
\[
AIC_k = -2 \hat{\ell}_k + 2 \# par,
\]
whereas the BIC index has expression 
\[
BIC_k = -2 \hat{\ell}_k + \log(n) \# par,
\]
where $\hat{\ell}_k$ denotes the maximum of the log-likelihood of the FMM with $k$ components and $\#par$ denotes the number of free parameters.
To select the optimal number of components, we estimate a series of models for increasing 
values of $k$,
and we select the one corresponding to the minimum value of these indexes.
The BIC  is usually preferred to the AIC as the latter
tends to overestimate the number of components \citep[][Chapter 6.9]{mcla:peel:00}.

Finally, on the basis of the estimation results, model-based clustering is performed with the Maximum-A-Posteriori (MAP) rule, consisting of assigning unit $i$ to the latent component 
corresponding to the maximum over $u$ of the posterior probability computed as in \eqref{eq:ppost}.
The selected  component for 
unit $i$ is dented by $\hat{u}_i$.
Note that, with missing data, we also perform a sort of multiple imputation that allows us  to predict the missing responses conditionally or unconditionally 
to the model component.
In the first case the predicted value is simply $\hat{\b y}_i=\E(\b Y_{i} | \b y_{i}^o, \hat{u}_i)$, while for the unconditional case it is computed as
\[
\tilde{\b y}_i=\sum_{u=1}^k\hat{z}_{iu}\E(\b Y_{i} | \b y_{i}^o, u),
\]
where the expected value is defined in \eqref{eq:cond_mean}. 
\section{Proposed hidden Markov model with missing data}\label{sec:HMmis} 
Considering the FMM outlined in the previous section, in the following we 
propose an HMM to deal with intermittent and monotone missing observations.
\subsection{Model formulation}
We denote by $\b Y_{it} = (Y_{i1t},\ldots,Y_{irt})\tr$ the vector of $r$ continuous response variables measured at time $t$, with $t=1,\ldots, T_i$.
Note that the number of  time occasions $T_i$ is 
specific for each individual  $i$, with $i=1,\ldots,n$. 
In this way, we also conceive unbalanced panels that, for example, are due to a different number of scheduled visits for every individual in a medical study. 
As already mentioned, in a longitudinal extension it is also important to account for dropout that gives rise to monotone informative missing data.
In this regard, we first introduce the indicator variable $D_{it}$ for the dropout, which assumes a value equal to 0 if  unit $i$ is still in the panel at occasion $t$ and equal to 1 if the same unit has dropped out.
It is worth noting that the event of dropout at time $t$, so that $D_{it}=1$, implies that  $D_{i,t+1}, \ldots, D_{iT_i}=1$.
Obviously, even if $D_{it}=0$, we may still have a missing observation at occasion $t$ due to the intermittent missing data pattern with all or some of the outcomes that are not observed.

The general HMM formulation assumes the existence of a latent process for each individual $i$, denoted by $\b U_i =(U_{i1},\ldots,U_{iT_i})'$, which affects the distribution of the response variables and  
is assumed to follow a first-order Markov chain with a certain number of states equal to $k+1$.
This model may account for dropout by adding an extra latent state, the $(k + 1)$-th, defined as an  absorbing state,  in the sense that once it has been reached, then it is not possible to move away from it. This proposal has been previously introduced in \cite{Montanari2018}. 

The HMM has two components: the measurement (sub)-model, concerning the conditional distribution of the response variables given the latent process, and the latent (sub)-model, concerning the distribution of the latent process.
Regarding the first component, we assume that the response vectors are conditionally independent given the hidden 
state and that, as in the FMM presented in Section \ref{sec:prel}, for the first $k$ states, the response vectors have Gaussian distribution with specific mean vector and variance-covariance matrix.
More precisely, we assume that
\[
\b Y_{it} | U_{it} = u, D_{it} = 0 \sim N(\b\mu_u, \b \Sigma), \quad \quad u=1,\ldots,k,
\]
where the means $\b\mu_u$, $u=1,\ldots,k$, are specific of each state, 
and the variance-covariance matrix $\b\Sigma$ is assumed to be constant across states under the assumption of homoschedasticity. 
Reducing the 
model complexity drives this choice, but homoschedasticity can be suitably relaxed if necessary. 
In addition,  
we assume that 
\[
P(D_{it} = d | U_{it} = u) = \left \{ \begin{array}{ll} 1 &  \text{with $d=0$ and $u = 1,\ldots, k$
}\\  \vspace{2mm}
& \:\:\text{\rm or $d=1$ and $u=k+1$}, \\ 
0 & \text{otherwise}. \end{array} \right. 
\]
In order to account for intermittent missing responses, we consider the partition ${\b Y}_{it} = (\b Y_{it}^o, \b Y_{it}^m)^\prime$, where $\b Y_{it}^o$ is 
the vector of the observed responses  and $\b Y_{it}^m$ is referred to the missing data.
As for the model illustrated 
in Section \ref{sec:prel}, %SP:,  
we consider a 
decomposition of the conditional mean vector and the variance-covariance matrix; see Expression \eqref{Eq:dec_mu_Si}.
Consequently, we have that
\[
\b Y_{it}^o 
| U_{it}= u, D_{it} = 0  \sim N(\b\mu_u^o,\b\Sigma^{oo}), \quad \quad u = 1,\ldots, k.
\]
Note that the distribution of 
$\b Y_{it}^{o}$
 given $U_{it} = k+1$ and $D_{it} = 1$ does not need to be defined.

Finally, the parameters of the latent model are the initial probabilities
\[
\pi_{u} = p(U_{i1} = u),\quad \quad u=1,\ldots k,
\]
and the transition probabilities 
\[
\pi_{u|\bu} = p(U_{it} = u| U_{i,t-1} = \bu), \quad \quad t = 2, \ldots,T_i, \; \:\ \bu, u = 1,\ldots,k, 
\]
where $u$ denotes a realization of $U_{it}$ and $\bu$ a realization of $U_{i,t-1}$. Note that the transition probabilities are assumed to be 
time homogeneous to reduce the number of free parameters, but even this assumption may be relaxed at  the occurrence. 
Moreover, given the interpretation of the latent state referred to 
as  dropout, the transition probabilities are suitably constrained as
\[
\pi_{k+1|k+1} = 1,\quad \pi_{u|k+1} = 0,  \quad  \quad u=1,\ldots,k.
\]
Therefore, once a subject reaches the $(k + 1)$-th latent state, he/she yields missing values until the end of the study. 
The interest in modeling this extra state is evident when the model includes individual covariates, as 
shown in the following.

In the present formulation, the manifest distribution is expressed with reference to the observed data represented by 
$\mathcal Y_i^o = \{\b y_{it}^o,  t=1,\ldots,T_i : d_{it}=0\}$, 
which is
the set of vectors $\b y_{it}^o$ observed  when $d_{it}=0$, for $i=1,\ldots,n$. 
Denoting by $\b d_i$ the  
vector of the observed indicator variables $D_{it}$ for individual $i$, we have that
\begin{eqnarray}\nonumber
f(\b d_{i}, \mathcal Y_i^o) &=& \sum_{\bl u_i}f(\mathcal Y_i^o | \b D_i =\b d_i,  \b U_i = \b u_i)   P(\b D_i  = \b d_i | \b U_i = \b u_i) P(\b U_i = \b u_i) \\
&=&\sum_{ \bl u_i}\left[\prod_{t=1}^{T_i} f(\b y_{it}^o| d_{it},u_{it}) \: p(d_{it}| u_{it}) \right]\left(\pi_{u_{i1}}\prod_{t=2}^{T_i}\pi_{u_{it}| u_{i,t-1}}\right).
\label{eq:manifest_miss}
\end{eqnarray}
As usual in dealing with HMMs, to efficiently compute this distribution, we can rely on a forward recursion \citep{baum:et:al:70,welch:2003}. 

The above formulation allows us to characterize the process generating informative dropout and to study the probability of transition from the estimated latent states to the dropout state. 
More in detail, starting from the EM algorithm illustrated in Section \ref{sec:EM_fin_mix}, we develop an inferential approach 
to obtain exact maximum likelihood estimates of model parameters under the MAR assumption for the intermittent missingness
and with informative dropout.
The resulting EM algorithm also includes suitable forward-backward recursions to perform the E-step \citep{baum:et:al:70,welch:2003}. 

\subsection{Inclusion of individual covariates}\label{sec:cov}
Longitudinal data allow for a  precise assessment of the effect 
of individual covariates
and this aspect is particularly relevant when these variables are related to a certain treatment as in the empirical illustration provided in Section \ref{sec:appl}.
In the HMM formulation, individual covariates may be included in the measurement model or in the latent model;  for a general review see \cite{bart:farc:penn:13} and  \citep{bart:farc:penn:14}. 
In the first case, the conditional distribution of the response variables given the latent states must be suitably parameterized.
In such a situation, the latent variables account for the unobserved heterogeneity that is allowed to be time-varying \citep{bart:farc:09}.
When covariates are included in the latent model,  the interest is in modeling the effect of covariates on the distribution of the latent process \citep{verm:lang:bock:99}.
This formulation is relevant  when the response variables measure an individual characteristic of interest represented by the latent variables.  

In this work, we consider the second formulation and we adopt a multinomial parameterization for the 
initial and transition Markov chain probabilities. 
More in detail, let $\b x_{it}$ denote the vector of individual covariates available at the $t$-th time occasion for individual $i$.
Now the initial and transition probabilities are individual specific and denoted by
$\pi_{iu} = p(U_{i1}=u|\b x_{i1})$, $u=1,\ldots,k$, and $\pi_{i,u|\bu} = p(U_{it}=u |U_{i,t-1}=\bu,\b x_{it})$, 
$t=2,\ldots,T$, $\bu,u=1,\ldots,k$, respectively.
We rely on the following parameterization 
\begin{eqnarray}  \label{eq:be}
 \log \frac{\pi_{iu}}{\pi_{i1}}=\beta_{0u}+\b x_{i1}\tr \b \beta_{1u},\quad u= 2,\ldots,k, 
\end{eqnarray}
for the  initial probabilities and on the following parametrization \begin{eqnarray}  \label{eq:Ga}
\log \frac{\pi_{i,u|\bu}}{\pi_{i,\bu|\bu}} = \gamma_{0\bu u}+\b x_{it}\tr
\b\gamma_{1\bu u},  \quad  \bu=1,\ldots,k,\; \:\ u = 1,\ldots,k+1,\; \:\ \bu\neq u, %SP: cambiato \bu con u
\end{eqnarray}
for the transition probabilities.
In the above expressions, $\b\beta_u=(\beta_{0u},\b\beta_{1u}\tr)\tr$ and $\b\gamma_{\bu u}=(\gamma_{0\bu u},\b\gamma_{1\bu u}\tr)\tr$ are parameter vectors to be estimated which are collected in the matrices $\b B$ and  $\b\Gamma$, respectively. 
Note that parameters in $\b B$ are not affected by the presence of the extra state since no unit is in the dropout state at the beginning of the study.
On the other hand, parameters in $\b\Gamma$ are properly constrained to avoid transitions from the 
%FP absorbing latent state.
latent absorbing state.

Finally, expression \eqref{eq:manifest_miss} for the manifest distribution is extended as
\begin{equation} 
f(\b d_{i}, \mathcal Y_i^o)  
= \sum_{ \bl u_i}\left[\prod_{t=1}^{T_i} f(\b y_{it}^o| d_{it},u_{it}) \: p(d_{it}| u_{it}) \right]\left(\pi_{i,u_{i1}}\prod_{t=2}^{T_i}\pi_{i,u_{it}| u_{i,t-1}}\right).\label{eq:manifestHMM_cov}
\end{equation}
\section{Model inference}\label{sec:inf}
In the following, we first illustrate the proposed inferential approach based on the maximization of the log-likelihood function.
Then, we outline the strategy for the initialization of the estimation algorithm.
Finally, we discuss issues related to the computation of the standard errors, 
selection of the number of states, and model-based dynamic clustering. 
\subsection{Maximum log-likelihood estimation with missing responses}\label{sec:MLE}
Assuming independence between sample units,
the log-likelihood  referred to the observed data is
\[
\ell (\b \theta) = \sum_{i=1}^n \log f(\b d_{i}, \mathcal Y_i^o). 
\]
In the above expression,  $\b \theta$ is the vector of all  model parameters and $f(\b d_{i}, \mathcal Y_i^o)$ 
is the manifest distribution of the observed responses defined in \eqref{eq:manifest_miss} and in \eqref{eq:manifestHMM_cov} with covariates. In order to estimate the parameters, we maximize $\ell (\b \theta)$ by an EM algorithm based on a {\em complete-data log-likelihood} that may be 
expressed as the sum of three components that are maximized separately:
\begin{eqnarray*}
\begin{split}
\ell^*(\b\th) =  \ell_1^*(\b\th) + \ell_2^*(\b\th) + \ell_3^*(\b\th),
\label{complik}
\end{split}
\end{eqnarray*}
where 
\begin{eqnarray*}
  \ell_1^*(\b\th) &=&   \sum_{i=1}^n \sum_{\substack{t=1 \\ (d_{it}=0)}}^{T_i} \sum_{u=1}^kz_{itu}\log f(\b y_{it}| D_{it}=0, u),\\
  \ell_2^*(\b\th) &=& \sum_{i=1}^n \sum_{u=1}^k z_{i1u}\log\pi_u,\\
  \ell_3^*(\b\th) &=& \sum_{i=1}^n \sum_{t=2}^{T_i} \sum_{\bu=1}^{k+1}\sum_{u=1}^{k+1}z_{it\bu u}\log \pi_{u|\bu}.
\end{eqnarray*}
In the above expressions, $z_{itu}=I(u_{it}=u)$ is an indicator variable equal to 1 if individual $i$ is in latent state $u$ at time $t$ and $z_{it\bu u}=z_{i,t-1,\bu}\;z_{itu}$ is the indicator variable for the transition from state $\bu$ to state $u$ of individual $i$ at time occasion $t$. 
When individual covariates are available, 
expressions for $ \ell_2^*(\b\th)$ and  $\ell_3^*(\b\th)$ are modified by substituting every $\pi_u$ and $\pi_{u|\bu}$ with $\pi_{iu}$ and $\pi_{i,u|\bu}$, respectively, which in turn are formulated as in expressions \eqref{eq:be} and \eqref{eq:Ga}.
Also note 
that in $\ell_1^*(\b\th)$, the sum over $t$ is computed only when $d_{it}=0$, whereas in $\ell_3^*(\b\th)$ 
the last two sums also involve the absorbing hidden 
state, $k+1$. 
More explicitly, the first component of the complete log-likelihood function may be written as
\begin{eqnarray*}
\begin{split}
\log f(\b y_{it}| D_{it}=0, u)=-\frac{1}{2}\log(|\b\Sigma|)-\frac{1}{2}(\b y_{it}-\b\mu_u)'\b\Sigma^{-1}(\b y_{it}-\b\mu_u).
\end{split}
\end{eqnarray*}
Therefore, we have that 
\[
\ell_1^*(\b\th) = \sum_{i=1}^n \bigg\{-\frac{T_i- d_{i+}}{2}\log(|\b\Sigma|) -\frac{1}{2}\sum_{\substack{t=1 \\ (d_{it}=0)}}^{T_i} \sum_{u=1}^kz_{itu}{\rm tr}\left[\b\Sigma^{-1}(\b y_{it}-\b\mu_u)(\b y_{it}-\b\mu_u)'\right]\bigg\},
\]
where $d_{i+}=\sum_{t=1}^{T_i}d_{it}$.

At the {\em E-step} of the EM algorithm we compute the posterior expected value of the indicator 
variables given the observed data and the current value of the parameters.
In particular, these expected values correspond to the following quantities
\begin{eqnarray}\label{eq:post1}
\hat{z}_{itu} & =& p(U_{it}=u |  \b y_{it}, \b d_i), \quad t=1,\ldots,T_i, \;  u = 1,\ldots,k+1\\\label{eq:post2}
\hat{z}_{it\bu u}&= &p(U_{it}=u, U_{i,t-1}=\bu |  \b y_{it},\b d_i), \quad t=2,\ldots,T_i, \; \:\ \bu,u = 1,\ldots,k+1,
\end{eqnarray}
computed by means of forward-backward  
recursions of  \cite{baum:et:al:70} and \cite{welch:2003}; for an illustration see \cite{bart:farc:penn:13}. 
We stress that when $d_{it}=1$, that is, when unit $i$ has dropped out at occasion $t$, we have $\hat{z}_{isu} = 0$, for $u = 1,\ldots,k$ and $s=t,\ldots, T_i$,  and $\hat{z}_{is,k+1} = 1$ for $s=t, \ldots, T_i$.
With individual covariates, the posterior probabilities 
in \eqref{eq:post1} and \eqref{eq:post2} are expressed 
by $\hat{z}_{itu}  = p(U_{it}=u |  \b y_{it}, \b d_i ,\b x_{it}) $  and $\hat{z}_{it\bu u}= p(U_{it}=u, U_{i,t-1}=\bu |  \b y_{it},\b d_i, \b x_{it})$, respectively. 
Furthermore, when $d_{it}=0$, the E-step  
also includes the computation of the following expected values resulting from the MAR assumption for the missing observations
\begin{eqnarray}\label{eq:Yimp}
\begin{split}
\E(\b Y_{it} | \b y_{it}^o, u) =\left(\begin{matrix}
\b  y_{it}^o \\
\b \mu_{u}^m + \b\Sigma^{mo} (\b\Sigma^{oo})^{-1} (\b y_{it}^o- \b \mu_{u}^o)
\end{matrix}\right),
\end{split}
\end{eqnarray}
\begin{eqnarray*}
\E\left[(\b Y_{it} - \b \mu_u)(\b Y_{it} - \b \mu_u)\tr| \b y_{it}^o, u\right]&=& \Var(\b Y_{it}| \b y_{it}^o) +  \E(\b Y_{it}| \b y_{it}^o, u)\E(\b Y_{it}| \b y_{it}^o, u)\tr\\
&&- \b \mu_u \E(\b Y_{it}| \b y_{it}^o, u)\tr- \E(\b Y_{it}| \b y_{it}^o, u)\b \mu_u\tr + \b \mu_u \b \mu_u \tr = \\
&=& \Var(\b Y_{it}| \b y_{it}^o) + [E(\b Y_{it}| \b y_{it}^o, u)- \b \mu_u] [E(\b Y_{it}| \b y_{it}^o, u)- \b \mu_u]\tr, 
\end{eqnarray*}
where 
\begin{eqnarray*}
\begin{split}
\Var(\b Y_{it}| \b y_{it}^o)  = \left(\begin{matrix}
\b O & \b O \\
\b O & \b \Sigma^{mm} - \b \Sigma^{mo}(\b\Sigma^{oo})^{-1}   \b \Sigma^{om}
\end{matrix}\right).
\end{split}
\end{eqnarray*}

At the  {\em M-step}  of the EM algorithm, we update the model parameters by considering the closed form solution for the means
\[
\b \mu_u = \frac{1}{\sum_{i=1}^n \sum_{\substack{t=1 \\ (d_{it}=0)}}^{T_i}\hat{z}_{itu}}\sum_{i=1}^n \sum_{\substack{t=1 \\ (d_{it}=0)}}^{T_i}\hat{z}_{itu} \E(\b Y_{it} | \b y_{it}^o, u), \quad u =1,\ldots,k,
\]
and 
we update  $\b \Sigma$ as 
\begin{eqnarray*}
\b \Sigma  &=& \frac{1}{\sum_{i=1}^n (T_i-d_{i+})}\sum_{i=1}^n\sum_{\substack{t=1 \\ (d_{it}=0)}}^{T_i}\sum_{u=1}^k \hat{z}_{itu}\bigg\{\Var(\b Y_{it}| \b y_{it}^o) + [\E(\b Y_{it}| \b y_{it}^o, u) -\b \mu_u]  [\E(\b Y_{it}| \b y_{it}^o, u) -\b \mu_u ]\tr  \bigg\} 
\end{eqnarray*}
that directly compare with Expressions
\eqref{eq:mu}  and \eqref{eq:Si}.

Finally, without individual covariates the initial and transition probabilities may be updated as
\begin{eqnarray*}
\pi_u &=& \frac{\sum_{i=1}^n \hat{z}_{i1u}}{n},\quad u = 1,\ldots,k,\\
\pi_{u|\bu} &=& \frac{\sum_{i=1}^n \sum_{t=2}^{T_i}\hat{z}_{it\bu u}}{\sum_{i=1}^n \sum_{t=2}^{T_i}\hat{z}_{i,t-1,u}},\quad \bu, u = 1,\ldots,k+1,
\end{eqnarray*}
whereas, with individual covariates, in order to update the latent model parameters we maximize the complete log-likelihood components $\ell_2^*(\b\th)$ and $\ell_3^*(\b\th)$, with respect to $\b B$ and $\b\Gamma$, by a Newton-Raphson algorithm.

\subsection{Algorithm initialization}\label{sec:ini}
As already mentioned for FMMs,
%FP inserito the
the initialization of the EM algorithm plays a central role as the model log-likelihood is typically multimodal.
This is a common problem in the estimation of discrete latent variable models implying that the EM algorithm may converge to one of the local modes that does not correspond to the global maximum.
In such a situation, a multi-start strategy, based both on a deterministic and a random starting rule, is necessary.
More in detail, the deterministic rule consists in computing the starting values of the parameters of the measurement model, $\b \mu_u$, $u=1,\ldots,k$, and $\b \Sigma$, on the basis of descriptive statistics (mean and covariance matrix) of the observed outcomes.
The starting values for the initial probabilities $\pi_u$ are chosen as $1/k$, for $u = 1,\ldots,k$, also including the constraint $\pi_{k+1}=0$.
For the transition probabilities we use $\pi_{u|\bu} = (h+1) / [h+(k+1)]$ when $u =\bu$ and  $\pi_{u|\bu} = 1/ [h+(k+1)]$
when $u \neq \bu$, for $u=1,\ldots, k+1$ and $\bu = 1,\ldots, k$, where $h$ is a suitable constant; for instance, in the
application illustrated in Section \ref{sec:appl} we use $h=9$. 
We also constrain the last row of the transition matrix to be $\pi_{u|k+1} = 0$ for $u=1,\ldots,k$ and $\pi_{k+1|k+1} = 1$.  
The random starting rule is based on values generated from a Gaussian distribution for the vectors $\b \mu_u$, $u=1,\ldots,k$, and on suitable normalized random numbers drawn from a uniform distribution  between 0 and 1 for both initial and transition probabilities. The starting values for the variance-covariance matrix are again chosen according to the covariance of the observed outcomes. 
The same rules may be 
suitably  adapted when individual covariates are included in the model; in this case, the initialization of the EM algorithm directly refers to  parameters in $\b B$ and $\b \Gamma$.

Overall, for a given $k$, the inference is  based on the solution corresponding to the largest value of the log-likelihood at convergence, which typically corresponds to the global maximum. 
The estimates obtained in this way are denoted by $\hat{\b\theta}$.
\subsection{Standard errors, model selection and clustering}\label{sec:se}
Once parameter estimates are computed for a given number of latent states $k$, and collected in $\hat{\b\theta}$, the 
corresponding standard errors may be obtained on the basis  of different methods. 
In this work, due to its ease of implementation and robustness of the corresponding results,  we mainly rely on a non-parametric bootstrap procedure \citep{davison:hynkley:1997}.
This  is performed by repeatedly sampling with replacement the 
data from the original sample, and fitting the proposed HMM with the selected number of states on these bootstrap samples.  
A drawback of this method is the high computational cost due to the need to fit the HMM for each resampled 
dataset.
A possible approach to reduce the computational burden is to select as starting values of the proposed EM algorithm the parameters obtained for the observed data. 

An alternative method to obtain standard errors is on the basis of the observed information matrix $J(\hat{\b\theta})$.
We compute this matrix through numerical methods as  proposed in \cite{bart:farc:09},  
that is,
as minus the numerical derivative of the score vector at convergence.
The score vector, in turn, is obtained as the first derivative of the expected value of the complete data log-likelihood, which is based on the expected frequencies $\hat{z}_{itu}$ and $\hat{z}_{it\bu u}$ corresponding to $\hat{\b\theta}$ \citep{oake:99}.
Accordingly, standard errors for the parameter estimates are obtained as the square root of the diagonal elements of the inverse of the observed information matrix $J(\hat{\b\theta})^{-1}$; 
see also \cite{bart:farc:15a}.

Concerning model selection, we rely on information criteria common to the finite-mixture literature \citep{mcla:peel:00} and, in particular, on the BIC, which outperforms the alternative information criteria as examined in \cite{bacc:pand:penn:14}.
Based on this selection approach, we estimate a series of models for increasing values of $k$, and we select the number of latent states corresponding to the minimum value of the BIC index.
However, as typically happens in applications to complex and high dimensional data, the BIC index may continue to decrease for each additional state added until a very large value of $k$. In such a situation, it is advisable to choose the value of $k$ that represents a suitable compromise between goodness-of-fit and interpretability of the resulting latent states as implemented, among others, in \cite{Montanari2018}.
It is also useful to visually display the values of the index against increasing $k$ so as to look for the ``elbow", that is, a change of slope in the curve suggesting the optimal number of states \citep{nylund2018}. 

Finally, once the number of states is selected, dynamic clustering is performed by assigning each unit to a latent state at each time occasion.
The EM algorithm directly provides the estimated posterior probabilities of $U_{it}$, as defined in \eqref{eq:post1}.
These probabilities can be directly used to perform {\em local decoding} so as to predict the latent states of each unit $i$ at each time occasion $t$.
To obtain the prediction of the latent trajectories of a unit across time, that is,  the most {\em a posteriori} 
 likely sequence of hidden 
 states, we also employ the so-called {\em global decoding}, which is based on an adaptation of the Viterbi algorithm \citep{vite:67}; see also \cite{juan:rabi:91}.

Even in this case, it is possible to perform multiple imputation of the missing responses conditionally or unconditionally 
to the predicted latent state. As in the context of FMMs, 
in the conditional case %SP:, 
the predicted value is simply $\hat{\b y}_{it}=\E(\b Y_{it} | \b y_{it}^o, \hat{u}_{it})$, where $\hat{u}_{it}$ are the predicted states.
The unconditional prediction of the missing responses is instead computed as
\[
\tilde{\b y}_{it}=\sum_{u=1}^k\hat{z}_{itu}\E(\b Y_{it} | \b y_{it}^o, u),
\]
where the expected value is defined as in \eqref{eq:Yimp}.

The functions used to perform maximum likelihood estimation of the proposed HMM with missing values, to compute standard errors, and to perform non-parametric bootstrap, are implemented by extending the functions of the \texttt{R} package {\tt LMest} \citep{bart:pand:penn:17} and are available at the GITHUB page ({\em GitHub link with the code will follow}). 
\section{Simulation Study}\label{sec:sim}
In the following, we illustrate the simulation design carried out to assess the performance of the proposed approach.
It is developed varying the sample size, the number of hidden states, and the assumed proportion of intermittent missing responses and informative dropout observations.
Here, we also aim at evaluating whether the proposed inferential approach allows us to identify the correct data generating process in terms of model parameters' recovery.
\subsection{Simulation design}
We randomly drew $B=250$ samples of size $n = 500, 1000$ from an 
HMM with a number of hidden states equal to $k=2,3$. We also considered a number of time occasions $T=5$ and $r=3$ continuous response variables, and a varying proportion of intermittent missing responses, that is, $p_{\rm miss} = 0.01, 0.05, 0.10, 0.25$. 

Regarding the measurement model, we considered the following values for the conditional means:
\begin{eqnarray*}
&& k=2: \b \mu_1 =  \begin{pmatrix*}[r] -2  \\ -2  \\ 0  \end{pmatrix*},\; \b\mu_2 = \begin{pmatrix*}[r] 0  \\ 2  \\ 2  \end{pmatrix*},\\
&& k=3: \b \mu_1 =  \begin{pmatrix*}[r] -2  \\ -2  \\ 0  \end{pmatrix*},\; \b\mu_2 = \begin{pmatrix*}[r] 0  \\ 0  \\ 0 \end{pmatrix*},\; \b\mu_3 = \begin{pmatrix*}[r] 0  \\ 2  \\ 2  \end{pmatrix*},
\end{eqnarray*}
and the following variance-covariance matrix
assumed 
constant across states:
\[
\b \Sigma = \begin{pmatrix} 1 & 0.5 & 0.5 \\
0.5 & 1 & 0.5 \\ 0.5 & 0.5 & 1 \end{pmatrix}.
\]
We considered equally likely hidden states at the first time period $\pi_u = 1/k$, with $u=1,\ldots,k$. 
Finally, we assumed a varying proportion of informative dropout, which was simulated by considering a transition matrix with increasing probabilities of moving towards the additional absorbing latent state, $k+1$, corresponding to the dropout, that is, $p_{\rm drop} = 0.01, 0.05, 0.10, 0.25$. This transition matrix is assumed to be time-homogeneous.
Overall, we 
considered  a total of 16 different scenarios, corresponding to the combination of $k=2,3$ latent states, $n=50,1000$ sample size, 
and $p_{\rm miss} = p_{\rm drop} = 0.01, 0.05, 0.10, 0.25$ for the proportion of intermittent missing responses and dropout.
\subsection{Results}
We assess the simulation results in terms of bias, standard deviation (sd), and root mean square error (rmse) of the parameter estimates. 
In particular, Table \ref{tab:1} reports the average, over the latent states and response variables, of the bias (in absolute value), sd, and rmse of the conditional mean vectors $\b\mu_u$, $u=1,\ldots,k$, under the different scenarios. 
Table \ref{tab:2} reports the same  estimation results, computed as the average over the different response variables, for the variance-covariance matrix $\b\Sigma$.
Finally, Tables \ref{tab:3} and \ref{tab:4} report the estimation results for the initial and transition probabilities, respectively.
  
Results highlight the ability of our approach in recovering the true data generating mechanism.
In particular, we observe that, regarding the estimation of all model parameters, the average bias, sd, and rmse are relatively small under all scenarios. 
Moreover, as expected, they tend to increase when considering a model with a higher number of hidden states. 
Moreover, the standard deviation and the root mean square error tend to decrease as the sample size increases. 
In general, model parameters are estimated with good accuracy, even in the presence of missing data.
The quality of results is only slightly affected by the presence of higher rates of intermittent missing data and/or dropout.

\begin{table}
  \centering
  \caption{\em Average, over the latent states and response variables, of the bias (in absolute value), standard deviation (sd), and root mean square error (rmse) of the conditional mean vectors $\b \mu_u$, $u=1,\ldots,k$}
    \begin{tabular}{c|c|rrrr}
    \toprule
   \multicolumn{1}{c}{}    &\multicolumn{1}{c}{}    & \multicolumn{2}{c}{$k=2$} & \multicolumn{2}{c}{$k=3$} \\
   \cline{3-6}
   \multicolumn{1}{c}{}         & \multicolumn{1}{c}{}  & \multicolumn{1}{c}{$n=500$} & \multicolumn{1}{c}{$n=1000$} & \multicolumn{1}{c}{$n=500$} & \multicolumn{1}{c}{$n=1000$} \\
          \midrule
  \multirow{3}[2]{*}{$p_{\rm miss} = p_{\rm drop} = 0.01$}& $\mid$ bias$\mid$ & 0.0016 & 0.0008 & 0.0016 & 0.0011 \\
   & sd & 0.0300 & 0.0206 & 0.0371 & 0.0265 \\
   & rmse & 0.0301 & 0.0206 & 0.0371 & 0.0266 \\
   \midrule
  \multirow{3}[2]{*}{$p_{\rm miss} = p_{\rm drop} = 0.05$}&    $\mid$bias$\mid$  &  0.0037 & 0.0010 & 0.0023 & 0.0016 \\
&   sd &  0.0324 & 0.0230 & 0.0399 & 0.0281 \\
 &  rmse & 0.0326 & 0.0230 & 0.0400 & 0.0281 \\
 \midrule
  \multirow{3}[2]{*}{$p_{\rm miss} = p_{\rm drop} = 0.10$}&    $\mid$ bias$\mid$  & 0.0022 & 0.0011 & 0.0021 & 0.0021 \\
&    sd & 0.0326 & 0.0240 & 0.0440 & 0.0312 \\
 & rmse& 0.0327 & 0.0240 & 0.0441 & 0.0313 \\
 \midrule
  \multirow{3}[2]{*}{$p_{\rm miss} = p_{\rm drop} = 0.25$}&    $\mid$ bias $\mid$ & 0.0058 & 0.0015 & 0.0019 & 0.0018 \\
&    sd & 0.0447 & 0.0304 & 0.0600 & 0.0415 \\
 &  rmse & 0.0451 & 0.0305 & 0.0600 & 0.0416 \\
 \bottomrule
    \end{tabular}%
  \label{tab:1}%
  \end{table}

\begin{table}[htbp]
  \centering
  \caption{\em Average, over the response variables, of the bias (in absolute value), standard deviation (sd), and root mean square error (rmse) of the variance and covariance matrix $\b\Sigma$}
    \begin{tabular}{c|c|rrrr}
\toprule
   \multicolumn{1}{r}{} & \multicolumn{1}{r}{} & \multicolumn{2}{c}{$k=2$} & \multicolumn{2}{c}{$k=3$} \\ 
   \cline{3-6}
   \multicolumn{1}{r}{} & \multicolumn{1}{r}{} & \multicolumn{1}{c}{$n=500$} & \multicolumn{1}{c}{$n=1000$} & \multicolumn{1}{c}{$n=500$} & \multicolumn{1}{c}{$n=1000$} \\
    \midrule
    \multirow{3}[2]{*}{$p_{\rm miss} = p_{\rm drop} = 0.01$} & $\mid$ bias$\mid$ &0.0020 & 0.0006 & 0.0012 & 0.0012 \\
          & sd &  0.0248 & 0.0176 & 0.0263 & 0.0187 \\
          & rmse & 0.0249 & 0.0176 & 0.0264 & 0.0187 \\
    \midrule
    \multirow{3}[2]{*}{$p_{\rm miss}= p_{\rm drop}  = 0.05$} & $\mid$ bias$\mid$ &  0.0017 & 0.0010 & 0.0019 & 0.0008 \\
          & sd & 0.0275 & 0.0202 & 0.0296 & 0.0206 \\
          & rmse & 0.0276 & 0.0203 & 0.0297 & 0.0206 \\
    \midrule
    \multirow{3}[2]{*}{$p_{\rm miss} = p_{\rm drop} = 0.10$} &$\mid$ bias$\mid$ & 0.0010 & 0.0014 & 0.0020 & 0.0017 \\
          & sd & 0.0309 & 0.0213 & 0.0312 & 0.0218 \\
          & rmse & 0.0309 & 0.0213 & 0.0313 & 0.0219 \\
    \midrule
    \multirow{3}[2]{*}{$p_{\rm miss} = p_{\rm drop} = 0.25$} &$\mid$ bias$\mid$& 0.0023 & 0.0014 & 0.0019 & 0.0017 \\
          & sd & 0.0395 & 0.0279 & 0.0422 & 0.0293 \\
          & rmse & 0.0396 & 0.0280 & 0.0423 & 0.0293 \\
    \bottomrule
    \end{tabular}%
  \label{tab:2}%
\end{table}

\begin{table}[htbp]
  \centering
  \caption{\em Average, over the latent states, of the bias (in absolute value), standard deviation (sd), and root mean square error (rmse) of the initial probabilities $\pi_u$, $u=1,\ldots,k$}
    \begin{tabular}{c|c|rrrr}
    \toprule
    \multicolumn{1}{r}{} & \multicolumn{1}{r}{} & \multicolumn{2}{c}{$k=2$} & \multicolumn{2}{c}{$k=3$} \\
   \cline{3-6}
    \multicolumn{1}{r}{} & \multicolumn{1}{r}{} & \multicolumn{1}{c}{$n=500$} & \multicolumn{1}{c}{$n=1000$} & \multicolumn{1}{c}{$n=500$} & \multicolumn{1}{c}{$n=1000$} \\
    \midrule
    \multirow{3}[2]{*}{$p_{\rm miss} = p_{\rm drop} = 0.01$} & $\mid$bias$\mid$ &0.0000 & 0.0002 & 0.0001 & 0.0004 \\
          & sd    &0.0153 & 0.0115 & 0.0173 & 0.0125 \\
          & rmse  & 0.0153 & 0.0115 & 0.0173 & 0.0125 \\
    \midrule
    \multirow{3}[2]{*}{$p_{\rm miss}= p_{\rm drop}  = 0.05$} & $\mid$bias$\mid$ & 0.0002 & 0.0002 & 0.0006 & 0.0006 \\
          & sd    &  0.0136 & 0.0105 & 0.0175 & 0.0124 \\
                    & rmse  & 0.0136 & 0.0106 & 0.0175 & 0.0125 \\
    \midrule
    \multirow{3}[2]{*}{$p_{\rm miss}= p_{\rm drop}  = 0.10$} & $\mid$bias$\mid$& 0.0001 & 0.0000 & 0.0009 & 0.0006 \\
          & sd    & 0.0168 & 0.0106 & 0.0176 & 0.0127 \\
          & rmse  & 0.0168 & 0.0106 & 0.0176 & 0.0127 \\
    \midrule
    \multirow{3}[2]{*}{$p_{\rm miss}= p_{\rm drop}  = 0.25$} & $\mid$bias$\mid$ & 0.0006 & 0.0001 & 0.0010 & 0.0004 \\
          & sd    & 0.0180 & 0.0116 & 0.0194 & 0.0143 \\
          & rmse  &  0.0180 & 0.0116 & 0.0195 & 0.0143 \\
    \bottomrule
    \end{tabular}%
  \label{tab:3}%
\end{table}

\begin{table}[htbp]
  \centering
  \caption{\em Average, over the latent states, of the bias (in absolute value), standard deviation (sd), and root mean square error (rmse) of the transition probabilities $\pi_{u|\bu}$, $u,\bu = 1,\ldots,k+1$}
    \begin{tabular}{c|c|rrrr}
    \toprule
    \multicolumn{1}{r}{} & \multicolumn{1}{r}{} & \multicolumn{2}{c}{$k=2$} & \multicolumn{2}{c}{$k=3$} \\
   \cline{3-6}
    \multicolumn{1}{r}{} & \multicolumn{1}{r}{} & \multicolumn{1}{c}{$n=500$} & \multicolumn{1}{c}{$n=1000$} & \multicolumn{1}{c}{$n=500$} & \multicolumn{1}{c}{$n=1000$} \\
    \midrule
    \multirow{3}[1]{*}{$p_{\rm miss} = p_{\rm drop} = 0.01$} & $\mid$bias$\mid$  &  0.0004 & 0.0002 & 0.0005 & 0.0003 \\
          & sd    & 0.0101 & 0.0072 & 0.0119 & 0.0085 \\
          & rmse  & 0.0052 & 0.0037 & 0.0075 & 0.0053 \\
          \midrule
    \multirow{3}[1]{*}{$p_{\rm miss}= p_{\rm drop}  = 0.05$} & $\mid$bias$\mid$  &0.0009 & 0.0001 & 0.0007 & 0.0004 \\
          & sd    & 0.0115 & 0.0081 & 0.0133 & 0.0095 \\
          & rmse  &  0.0069 & 0.0048 & 0.0092 & 0.0066 \\
          \midrule
    \multirow{3}[1]{*}{$p_{\rm miss} = p_{\rm drop} = 0.10$} & $\mid$bias$\mid$  & 0.0009 & 0.0004 & 0.0005 & 0.0005 \\
          & sd    & 0.0120 & 0.0094 & 0.0150 & 0.0105 \\
          & rmse  &0.0078 & 0.0058 & 0.0110 & 0.0077 \\
          \midrule
     \multirow{3}[1]{*}{$p_{\rm miss} = p_{\rm drop} = 0.25$} & $\mid$bias$\mid$  & 0.0009 & 0.0005 & 0.0011 & 0.0007 \\
          & sd    & 0.0161 & 0.0117 & 0.0190 & 0.0137 \\
                    & rmse  &  0.0109 & 0.0080 & 0.0150 & 0.0108 \\
    \bottomrule
    \end{tabular}%
  \label{tab:4}%
\end{table}

In order to evaluate how increasing frequencies of informative dropout affect the results, we also report in Table \ref{tab:5} the average, over the random samples, of the estimated transition matrix under the scenario with $k=3$ latent states. 
Note that we denote with ${drop}$ the additional  state corresponding to the absorbing/dropout state.
\begin{table}[htbp]
  \centering
 \caption{\em Averaged estimated transition matrix under the HMM  with $k=3$ latent states}
    \begin{tabular}{cl|rrrr|rrrr}
      \toprule  
          &   \multicolumn{1}{c}{}    & \multicolumn{4}{c}{$n=500$} & \multicolumn{4}{c}{$n=1000$} \\
   \cline{3-10}
          &   \multicolumn{1}{c}{}     & \multicolumn{1}{l}{$u=1$} & \multicolumn{1}{l}{$u=2$} & \multicolumn{1}{l}{$u=3$} & \multicolumn{1}{l}{$drop$} & \multicolumn{1}{l}{$u=1$} & \multicolumn{1}{l}{$u=2$} & \multicolumn{1}{l}{$u=3$} & \multicolumn{1}{l}{$drop$} \\
          \midrule
    \multirow{4}[0]{*}{$p_{\rm miss} = p_{\rm drop} = 0.01$} & $u=1$   & 0.889 & 0.092 & 0.010 & 0.010 & 0.890 & 0.090 & 0.010 & 0.010 \\
          & $u=2$   & 0.080 & 0.830 & 0.080 & 0.010 & 0.080 & 0.829 & 0.081 & 0.010 \\
          & $u=3$   & 0.010 & 0.088 & 0.892 & 0.010 & 0.010 & 0.090 & 0.891 & 0.010 \\
          & $drop$ & 0.000 & 0.000 & 0.000 & 1.000 & 0.000 & 0.000 & 0.000 & 1.000 \\
          \midrule
    \multirow{4}[0]{*}{$p_{\rm miss}= p_{\rm drop}  = 0.05$} & $u=1$   & 0.847 & 0.092 & 0.011 & 0.050 & 0.849 & 0.091 & 0.010 & 0.050 \\
          & $u=2$   & 0.081 & 0.791 & 0.078 & 0.050 & 0.081 & 0.791 & 0.079 & 0.050 \\
          & $u=3$   & 0.010 & 0.090 & 0.849 & 0.051 & 0.010 & 0.090 & 0.850 & 0.050 \\
          & $drop$   & 0.000 & 0.000 & 0.000 & 1.000 & 0.000 & 0.000 & 0.000 & 1.000 \\
                \midrule
    \multirow{4}[0]{*}{$p_{\rm miss}= p_{\rm drop}  = 0.1$} & $u=1$   & 0.801 & 0.091 & 0.010 & 0.098 & 0.801 & 0.091 & 0.009 & 0.099 \\
          & $u=2$   & 0.081 & 0.739 & 0.081 & 0.099 & 0.080 & 0.741 & 0.080 & 0.099 \\
          & $u=3$  & 0.010 & 0.091 & 0.800 & 0.100 & 0.010 & 0.090 & 0.801 & 0.099 \\
          & $drop$   & 0.000 & 0.000 & 0.000 & 1.000 & 0.000 & 0.000 & 0.000 & 1.000 \\
                          \midrule
    \multirow{4}[0]{*}{$p_{\rm miss}= p_{\rm drop}  = 0.25$} & $u=1$   & 0.648 & 0.091 & 0.011 & 0.251 & 0.648 & 0.091 & 0.010 & 0.251 \\
          & $u=2$   & 0.081 & 0.587 & 0.079 & 0.253 & 0.082 & 0.589 & 0.079 & 0.250 \\
          & $u=3$  & 0.011 & 0.090 & 0.652 & 0.247 & 0.010 & 0.090 & 0.651 & 0.249 \\
          & $drop$   & 0.000 & 0.000 & 0.000 & 1.000 & 0.000 & 0.000 & 0.000 & 1.000 \\
          \bottomrule
    \end{tabular}
  \label{tab:5}
\end{table}
From these results, we may observe that the probability of moving toward the dropout state is appropriately estimated as the dropout proportion  
increases, regardless of the sample size.

\clearpage
\section{Application}\label{sec:appl}
In the following we describe the historical data on primary biliary cholangitis, and then we illustrate the results obtained through the application of the proposed model.

\subsection{Data description}
The data we use to illustrate our proposal come from 
biochemical measurements collected prospectively by the Mayo Clinic from January 1974 to May 1984 \citep{murt94}
\footnote{The data in different formats are available from the R package \texttt{JM} \citep{rizo:10} and on-line in the datasets archive at the website: \url{http://lib.stat.cmu.edu/datasets/pbcseq}}. They derive from  a randomized control trial related to the primary biliary cholangitis (or cirrhosis, PBC), which  is a liver disease implying inflammatory destruction of the bile ducts and eventually leads to cirrhosis of the liver \citep{dick:89}.
It is a chronic disease of unknown causes with a prevalence of about 50-cases-per-million population.

Data are referred to $n= 312$ patients,  some of which (158)  were randomized to D-penicillamine and some others (154) with placebo. 
The original clinical protocol for these patients specified visits at 6 months, 1 year, and annually after that. However,  the actual follow-up times varied considerably around the scheduled visits.
Therefore, we considered  time occasions at 6 months from the baseline, thus accounting for missing observations, missing visits, and dropout in a period of 29 time occasions. 
These data  have been   frequently analyzed through the  Cox hazard model \citep{cox:72} and,  more recently, by joint models \citep{rizo:12,bartolucci2015discrete}.
In these previous works, despite the immunosuppressive properties of D-penicillamine, no relevant differences were observed between the distribution of treated and untreated patients' survival times.
In this context it is often of interest to account for a  multivariate analysis of the longitudinally collected measurements for the  diagnosing of  liver diseases.
Moreover, as remarked in \cite{rizo:12}, physicians are interested in measuring the joint association of the levels of biomarkers with the risk of death. 
 
In the present application, we considered  the following  biochemical variables: {\em bilirubin}, {\em cholesterol},  {\em albumin},  {\em platelets},  {\em prothrombin},  {\em alkaline}, and {\em transaminase}. 
There are some atypical observations or outliers in the data, and for this reason, we chose  to consider the natural logarithm of the biomarkers. 
In addition to drug use, we  considered gender and age as covariates.  
We noticed that the sample is not balanced according to gender since 88\% of the sample are women with an average age of 50. 
Age is considered a time-varying  covariate, and along with gender, we investigate how it is associated with the risk of dropout.
Descriptive statistics of the responses, dropout rates for treated and untreated patients, and additional details on covariates along with a figure of the observed values for each patient on each response at every time occasion are provided in the Supporting Information (SI). 
\subsection{Results}
The proposed HMM allows us to jointly account for the missing mechanism and the complex censoring  mechanism \citep{rub:76} involved in the PBC study. 
First, we estimated the HMM without covariates and  with homogeneous transition probabilities in order to select a suitable number of latent states.
The  initialization strategy illustrated in Section \ref{sec:ini} is adopted for the EM algorithm. 
In particular, after a deterministic initialization, a number of random initializations equal to $5\times k$ is considered, with $k$ denoting the number of latent  states ranging from 1 to 8. 
Table  \ref{tab:bic}  reports  the results of the fitting procedure.
We notice that  the decrease in the BIC index 
obtained with the model that has more than five latent states is relatively lower than that obtained with fewer states.
Therefore, as discussed in Section \ref{sec:se}, for the parsimony principle, we selected the HMM with $k=5$ latent states. Then, we included the individual covariates and we estimated the HMM proposed in Section \ref{sec:cov} keeping the number of states fixed. %SP: reinserita frase altrimenti non si capiva dove venivano inserite le covariate

\begin{table}[htbp]
\centering
\renewcommand{\arraystretch}{0.8}
\caption{\em{Results from the fitting of the multivariate  HMMs for an  increasing number of hidden states ($k$)}}
\vspace{-0.3cm}
\begin{tabular}{lrrrrr}
\midrule
     \multicolumn{1}{c}{$k$} & \multicolumn{1}{c}{$\hat{\ell}$} & \multicolumn{1}{c}{$\#$par} & \multicolumn{1}{c}{$BIC_k$} & \multicolumn{1}{c}{$AIC_k$}  \\ 
\midrule
1 & -60,510.60 & 35 & 121,222.21 & 121,091.20 \\ 
  2 & -36,79.84 & 45 & 7,618.11 & 7,449.67 \\ 
  3 & -32,42.58 & 57 & 6,812.51 & 6,599.16 \\ 
  4 & -2,879.04 & 71 & 6,165.84 & 5,900.09 \\ 
  5 & -2,561.74 & 87 & 5,623.13 & 5,297.49 \\ 
  6 & -2,404.29 & 105 & 5,411.60 & 5,018.58 \\ 
  7 & -2,314.55 & 125 & 5,346.98 & 4,879.11 \\ 
  8 & -2,215.85 & 147 & 5,275.92 & 4,725.70 \\  \bottomrule
\end{tabular}\label{tab:bic}
\end{table}
Table  \ref{tab:means} shows the estimated conditional means, $\b \mu_u$, 
$u=1,\ldots,5$, of the biomarkers (in logarithm) of the model with covariates.
We notice that the 3rd and the 5th states include patients in the worst health conditions since they show the highest {\em bilirubin} values.
However, the 5th state is also characterized by the lowest average of {\em albumin}, the highest average of {\em prothrombin}, and by high values of {\em alkaline} and {\em transaminase}. %FB: se la bilirubina è la grandezza principale (da quanto capisco) non conviene riordinare gli stati secondo questa grandezza? se troppo complicato lasciamo perdere semmai lo vediamo meglio più avanti
%SP: non so Fulvia cosa ne pensa, secondo me potrebbe essere complicato rifare tutti i commenti
%FP occorre cambiare: tutte le tabelle dell'applicazione e le figure e le ultime due tabelle del SI e rifare i commenti in base al nuovo ordine: abbiamo già presentato il paper così magari potevamo decidere questo prima, in ogni caso ci vuole e ora io non riesco se volete posso farlo la prossima settimana come preferisci Silvia
%SP: per me ok così per ora
The 1st, 2nd, and  4th states are referred to patients in quite relatively good conditions 
with respect to the values of the biomarkers.
\begin{table}%[h!]
\centering
\renewcommand{\arraystretch}{0.8}
\caption{\em{Estimated conditional means $\b\mu_u$, $u=1,\ldots,k$ of the biomarkers (in logarithm), 
under the  HMM with $k$ = 5 hidden states}}\label{tab:means}
\vspace{-0.3cm}
\begin{tabular}{lrrrrrrr}
\toprule
& \multicolumn{1}{c}{1} & \multicolumn{1}{c}{2} & \multicolumn{1}{c}{3} & \multicolumn{1}{c}{4} & \multicolumn{1}{c}{5}  \\ 
\midrule
 {\em Bilirubin} & 0.136 & 0.865 & 2.020 & -0.432 & 2.411 \\ 
  {\em Cholesterol}   & 5.783 & 5.496 & 6.146 & 5.508 & 5.415 \\ 
  {\em Albumin}  & 1.270 & 1.137 & 1.177 & 1.279 & 0.940 \\ 
  {\em Platelets}   & 5.556 & 4.776 & 5.525 & 5.477 & 5.010 \\ 
  {\em Prothrombin}  & 2.339 & 2.441 & 2.396 & 2.363 & 2.578 \\ 
  {\em Alkaline}  & 7.270 & 6.824 & 7.611 & 6.430 & 7.033 \\ 
  {\em Transaminase}  & 4.769 & 4.664 & 5.163 & 4.086 & 5.070 \\ 
\bottomrule
\end{tabular}
\end{table}

Table \ref{tab:var} shows the estimated variances and covariances and the partial correlations, from which we observe that    {\em bilirubin} and   {\em prothrombin} have a  positive correlation given all the remaining biochemical measurements, as well as   {\em alkaline} and   {\em transaminase} (0.256) and   {\em alkaline} and   {\em cholesterol} (0.242).
We also have  a negative partial correlation among   {\em albumin} and    {\em prothrombin}  (-0.148).
\begin{table}%[h!]
\centering
\renewcommand{\arraystretch}{0.8}
\caption{\em{Estimated variance-covariances  (lower part, in bold estimated variances) and the estimated partial correlations (upper part) under %SP:of  
the 
HMM with $k=5$ hidden states}}
\begin{tabular}{lrrrrrrr}
\toprule
 Responses  & \multicolumn{1}{c}{\em{Biril}} & \multicolumn{1}{c}{\em{Chol}} & \multicolumn{1}{c}{\em{Albu}} & \multicolumn{1}{c}{\em{Plat}} & \multicolumn{1}{c}{ \em{Proth}}  & \multicolumn{1}{c}{\em{Alka}}  & \multicolumn{1}{c}{\em{Tran}}  \\ 
\midrule
     {\em Bilirubin} & \bf{0.270} & 0.092 & -0.130 & -0.108 & 0.257 & -0.026 & 0.239 \\ 
    {\em Cholesterol} & 0.027 & \bf{0.087} &  0.066 & 0.156 & -0.112 & 0.242 & 0.023 \\ 
    {\em Albumin} & 0.000 & -0.001 & \bf{0.017} & -0.078 & -0.148 & -0.074 & 0.054 \\ 
    {\em Platelets} & -0.016 & 0.014 & -0.002 & \bf{0.118} & -0.141 & 0.126 & 0.008 \\ 
    {\em Prothrombin}  & 0.005 & -0.001 & -0.001 & -0.002 & \bf{0.008} & 0.021 & -0.025  \\ 
    {\em Alkaline} & 0.039 & 0.038 & -0.005 & 0.026 & 0.002 & \bf{0.246} &  0.256\\ 
    {\em Transaminase} & 0.062 & 0.015 & -0.000 & -0.005 & 0.001 & 0.062 & \bf{0.161} \\   
\bottomrule
\end{tabular}\label{tab:var}
\end{table}

Table \ref{tab:initrans} reports the estimated averaged initial and transition probabilities, computed with respect to all patients in the sample,
 where we notice that  at the baseline, the 1st state is the most likely since  44\% of patients are in this state and 21\%, 19\% and 12\% are in the 3rd, 4th, and 2nd state, respectively. The 5th state, which includes patients with worse health conditions, has the 4.2\% of patients.
According to the estimated averaged 
transition probabilities, the most persistent state is the 2nd, whereas  
the state with the highest probability towards dropout  is the 5th ($\hat{\pi}_{drop|5} = 0.265$) followed by the 3rd state  ($\hat{\pi}_{drop|3} = 0.022$). 
The 1st,  2nd, and 4th  states  may lead to dropout as well, but with lower probabilities. Patients in the 3rd latent state have a probability of moving to the 5th state equal to $\hat{\pi}_{5|3}  = 0.094$ that is the highest estimated probability out of the main diagonal  of the transition matrix, excluding those of the dropout state. 
These results also show 
that  higher values of serum bilirubin and prothrombin and lower albumin values are strongly related to the risk for death. 
\begin{table}%[ht]
\centering
\renewcommand{\arraystretch}{0.8}
\caption{{\em Estimated initial and transition probabilities  under the HMM with $k$ = 5 hidden states}}
\vspace{-0.3cm}
\begin{tabular}{lrrrrrrrr}
 \toprule
 \renewcommand{\arraystretch}{3}
&  \multicolumn{6}{c}{$u$}\\
\cline{2-7}
 &  \multicolumn{1}{c}{1} & \multicolumn{1}{c}{2} & \multicolumn{1}{c}{3} & \multicolumn{1}{c}{4} &\multicolumn{1}{c}{ 5} & \multicolumn{1}{c}{$drop$}  \\ 
\midrule
\: $\hat{\pi}_u$ &  0.438 & 0.121&  0.208 & 0.191 & 0.042 & 0.000 \\
\midrule
   $\hat{\pi}_{u|1}$ & 0.868 & 0.056 & 0.020 & 0.040 & 0.012 & 0.004 \\ 
   $\hat{\pi}_{u|2}$ & 0.000 & 0.922 & 0.004 & 0.000 & 0.066 & 0.008 \\ 
   $\hat{\pi}_{u|3}$ & 0.002 & 0.008 & 0.864 & 0.010 & 0.094 & 0.022 \\ 
   $\hat{\pi}_{u|4}$ & 0.021 & 0.003 & 0.000 & 0.939 & 0.036 & 0.001 \\ 
   $\hat{\pi}_{u|5}$ & 0.000 & 0.014 & 0.014 & 0.053 & 0.654 & 0.265 \\ 
    $\hat{\pi}_{u|drop}$ & 0.000 & 0.000 & 0.000 & 0.000 & 0.000 & 1.000 \\ 
  \bottomrule
\end{tabular}\label{tab:initrans}
\end{table}

Table \ref{tab:inicov}  provides the estimated regression parameters for the initial probabilities where the statistical significance of the coefficients is established according to the estimated standard errors obtained with the non-parametric bootstrap.
These standard errors are reported in Tables 2  of the SI, whereas Table 3 of the SI  also  shows the estimated standard errors obtained  by  using the information matrix as explained in Section \ref{sec:se}.
\begin{table}%[htb]
\centering
\renewcommand{\arraystretch}{0.8}
\caption{\em  Estimates of the logit regression parameters  of the initial probability %of  belonging  %SP:to belong 
%FP lascrei to belong 
to belong  to the other latent states with respect to the 
1st state under the HMM with $k=5$ hidden states (significant $^\dag$at 10\%, $^*$at 5\%, $^{**}$at 1\%)}\label{tab:inicov}
\vspace{-0.3cm}
\begin{tabular}{lrrrrrrr}
\toprule
Effect  &  \multicolumn{1}{c}{$\hat{\beta}_{12}$} & \multicolumn{1}{c}{$\hat{\beta}_{13}$} & \multicolumn{1}{c}{$\hat{\beta}_{14}$} & \multicolumn{1}{c}{$\hat{\beta}_{15}$} \\
\midrule	
Intercept & -5.028$^{**}$ & -0.356 & -4.403$^{*}$ & -13.506$^{**}$ \\ 
Drug & 0.544 \:\:\ & -0.322 & 0.341\:\:\  & -0.297\:\:\:\:\  \\ 
Female & -0.124\:\:\:\  & -0.412 & 1.204\:\:\  & 6.634$^{**}$\:\ \\ 
Age & 0.068$^{**}$ & 0.002 & 0.046$^{*}$ & 0.093$^{*}$\:\:\  \\ 		
\bottomrule
\end{tabular}
\end{table}
The estimated gender log-odds in Table \ref{tab:inicov} relative to the 5th state is positive and significant indicating that the probability of being in the 5th state at the beginning of the study is higher for females 
with respect to males.
The log-odds related to age are positive, indicating that, at the baseline, older patients generally tend to belong  to the other states with respect to the 1st. 
\begin{table}%[htb]
\centering
\renewcommand{\arraystretch}{0.8}
\caption{\em  Estimates of the logit regression parameters  of the transition probabilities  under the HMM with $k=5$ hidden states  (significant $^\dag$at 10\%, $^*$at 5\%, $^{**}$at 1\%)} \label{tab:trancov}
\begin{tabular}{lrrrrrrr}
\toprule
Effect &
 \multicolumn1c{$\hat{\gamma}_{112}$} & \multicolumn1c{$\hat{\gamma}_{113}$} &\multicolumn1c{$\hat{\gamma}_{114}$} & \multicolumn1c{$\hat{\gamma}_{115}$}   & \multicolumn1c{$\hat{\gamma}_{11drop}$}  \\ 
 \midrule
Intercept & -4.015$^{**}$ & -2.686\:\:\:\ & -5.569\:\:\:\ & -20.658$^{**}$ & -10.749$^{**}$ \\ 
  Drug & 0.601\:\:\:\ & 0.266\:\:\:\ & 1.319\:\:\:\ & 8.223$^{**}$ & 1.193\:\:\:\ \\ 
  Female & -1.150\:\:\:\ & -0.347\:\:\:\ & -0.420\:\:\:\ & -10.402$^{**}$ & 5.329$^{**}$ \\ 
  Age & 0.033\:\:\:\ & -0.015\:\:\:\ & 0.035\:\:\:\ & 0.165\:\:\:\ & -0.012\:\:\:\ \\ 	\midrule
Effect  &
 \multicolumn1c{$\hat{\gamma}_{121}$} & \multicolumn1c{$\hat{\gamma}_{123}$} &\multicolumn1c{$\hat{\gamma}_{124}$} & \multicolumn1c{$\hat{\gamma}_{125}$}  & \multicolumn1c{$\hat{\gamma}_{12drop}$}   \\ 
 \midrule
Intercept & -36.680$^{**}$ & 10.020$^{**}$ & -20.780$^{**}$ & -4.705$^{**}$ & -7.047\:\:\:\ \\ 
  Drug & 1.207$^\dag$\:\:\ & 21.313$^{**}$ & -4.293$^{**}$ & -0.040\:\:\:\ & 1.075\:\:\:\ \\ 
  Female & -6.604$^{**}$ & -25.621$^{**}$ & 2.015$^{*}$\:\:\ & 0.117\:\:\:\ & -1.430\:\:\:\ \\ 
  Age & 0.231$^{**}$ & -0.757$^{**}$ & -0.017\:\:\:\ & 0.034\:\:\:\ & 0.045\:\:\:\ \\ 	\midrule
Effect  &
 \multicolumn1c{$\hat{\gamma}_{131}$} & \multicolumn1c{$\hat{\gamma}_{132}$} &\multicolumn1c{$\hat{\gamma}_{134}$} & \multicolumn1c{$\hat{\gamma}_{135}$}  & \multicolumn1c{$\hat{\gamma}_{13drop}$}    \\ 
 \midrule
Intercept & -7.228\:\:\:\ & -5.091\:\:\:\ & -24.962$^{\dag}$ & -3.817$^{\dag}$ & -8.306\:\:\:\ \\ 
  Drug & -6.792$^{**}$ & 2.355\:\:\:\ & 4.426\:\:\:\ & -0.894\:\:\:\ & 1.203\:\:\:\ \\ 
  Female & 4.579$^{\dag}$\:\:\ & 9.356\:\:\:\ & 5.484\:\:\:\ & -0.674\:\:\:\ & 1.127\:\:\:\ \\ 
  Age & -0.050\:\:\:\ & -0.227\:\:\:\ & 0.173\:\:\:\ & 0.044$^{\dag}$\:\:\ & 0.048\:\:\:\ \\ \midrule
Effect  &
 \multicolumn1c{$\hat{\gamma}_{141}$} & \multicolumn1c{$\hat{\gamma}_{142}$} &\multicolumn1c{$\hat{\gamma}_{143}$} & \multicolumn1c{$\hat{\gamma}_{145}$}  & \multicolumn1c{$\hat{\gamma}_{14drop}$}   \\ 
\midrule
Intercept & 20.506\:\:\:\ & -17.921$^{**}$ & -31.507$^{**}$ & -20.653\:\:\:\ & -20.304$^{**}$ \\ 
  Drug & -17.895$^{**}$ & 8.795$^{**}$ & -4.017$^{**}$ & -1.745\:\:\:\ & 7.761$^{**}$ \\ 
  Female & -20.685$^{**}$ & 6.639$^{**}$ & -3.181$^{**}$ & -6.747\:\:\:\ & 5.205$^{\dag}$\:\:\ \\ 
  Age & -0.316\:\:\:\ & -0.044\:\:\:\ & 0.176\:\:\:\ & 0.309$^{**}$ & 0.023\:\:\:\ \\\midrule 
Effect  &
 \multicolumn1c{$\hat{\gamma}_{151}$} & \multicolumn1c{$\hat{\gamma}_{152}$} &\multicolumn1c{$\hat{\gamma}_{153}$} & \multicolumn1c{$\hat{\gamma}_{154}$}  & \multicolumn1c{$\hat{\gamma}_{15drop}$}   \\ 
\midrule
Intercept & -6.887$^{**}$ & 9.574$^{**}$ & 3.689\:\:\:\ & -5.866$^{\dag}$\:\:\ & -0.825\:\:\:\ \\ 
  Drug & -3.753$^{**}$ & -4.542\:\:\:\ & 5.982\:\:\:\ & -1.389\:\:\:\ & -0.595\:\:\:\ \\ 
  Female & 4.468$^{**}$ & 23.414$^{**}$ & -0.674\:\:\:\ & 6.848$^{**}$ & 0.400\:\:\:\ \\ 
  Age & -0.289$^{**}$ & -0.892$^{**}$ & -0.261\:\:\:\ & -0.051\:\:\:\ & -0.001\:\:\:\ \\ 
\bottomrule
\end{tabular}
\end{table}
The estimated parameters in Table \ref{tab:trancov} refer to the coefficients affecting the transition from level $\bar{u}$ to level $u$ of the latent process and, for example, we notice that the first column of the last panel of the table contains the parameter estimates measuring the influence of each covariate on the transition from the 5th state, corresponding to the worst health conditions,  to the 1st state, 
corresponding to the best health conditions.
The influence of  gender is positive,  indicating that this transition probability is higher for females than 
males.
On the other hand, age negatively affects the same transition.
Another interpretation of the effects of covariates can be retrieved by looking at the 
estimated 
 initial and transition probabilities defined by categories of patients. 
Based on these average estimated values, we can study if the disease evolution is 
different  according to drug use, gender, and age. 
\begin{table}%[ht]
\centering
\renewcommand{\arraystretch}{0.8}
\caption{\em Average  %SP: average of 
of the estimated initial and transition probabilities with respect to  
the drug use (upper panel)  or not use (bottom panel)} \label{tab:trans_drug}
\vspace{-0.1cm}
\begin{tabular}{lrrrrrrrr}
 \toprule
&  \multicolumn{6}{c}{$u$}\\
\cline{2-7}
 &  \multicolumn{1}{c}{1} & \multicolumn{1}{c}{2} & \multicolumn{1}{c}{3} & \multicolumn{1}{c}{4} &\multicolumn{1}{c}{ 5} & \multicolumn{1}{c}{$drop$}  \\ 
  \midrule
  \: $\hat{\pi}_u$ &  
   0.419 & 0.155 & 0.169 & 0.219 & 0.038 & 0.000 \\    \midrule
   $\hat{\pi}_{u|1}$ & 0.818 & 0.070 & 0.022 & 0.062 & 0.023 & 0.005 \\ 
   $\hat{\pi}_{u|2}$   & 0.000 & 0.912 & 0.009 & 0.000 & 0.067 & 0.012 \\ 
  $\hat{\pi}_{u|3}$ & 0.000 & 0.014 & 0.870 & 0.020 & 0.062 & 0.034 \\ 
   $\hat{\pi}_{u|4}$ & 0.000 & 0.006 & 0.000 & 0.957 & 0.035 & 0.002 \\ 
   $\hat{\pi}_{u|5}$  & 0.000 & 0.003 & 0.027 & 0.023 & 0.729 & 0.218 \\ 
  $\hat{\pi}_{u|drop}$ & 0.000 & 0.000 & 0.000 & 0.000 & 0.000 & 1.000 \\
\midrule \renewcommand{\arraystretch}{3}
 &  \multicolumn{1}{c}{1} & \multicolumn{1}{c}{2} & \multicolumn{1}{c}{3} & \multicolumn{1}{c}{4} &\multicolumn{1}{c}{ 5} & \multicolumn{1}{c}{$drop$}  \\   
 \midrule 
 \: $\hat{\pi}_u$ &  
 0.513 & 0.071 & 0.236 & 0.158 & 0.022 & 0.000   \\  \midrule
$\hat{\pi}_{u|1}$ & 0.905 & 0.039 & 0.023 & 0.028 & 0.001 & 0.004 \\ 
   $\hat{\pi}_{u|2}$  & 0.000 & 0.938 & 0.009 & 0.000 & 0.049 & 0.004 \\ 
  $\hat{\pi}_{u|3}$ & 0.003 & 0.015 & 0.903 & 0.001 & 0.064 & 0.014 \\ 
  $\hat{\pi}_{u|4}$ & 0.017 & 0.004 & 0.000 & 0.978 & 0.000 & 0.001 \\ 
  $\hat{\pi}_{u|5}$ & 0.000 & 0.029 & 0.027 & 0.075 & 0.609 & 0.260 \\ 
  $\hat{\pi}_{u|drop}$ & 0.000 & 0.000 & 0.000 & 0.000 & 0.000 & 1.000 \\ 
 \bottomrule
\end{tabular}
\end{table}
\begin{table}%[ht]
\centering
\renewcommand{\arraystretch}{0.8}
\caption{\em Average of the estimated  initial  and  transition probabilities  
with respect to gender: 
female (upper panel)  or male (bottom panel)} \label{tab:trans_gen}
\vspace{-0.1cm}
\begin{tabular}{lrrrrrrrr}
 \toprule
&  \multicolumn{6}{c}{$u$}\\
\cline{2-7}
 &  \multicolumn{1}{c}{1} & \multicolumn{1}{c}{2} & \multicolumn{1}{c}{3} & \multicolumn{1}{c}{4} &\multicolumn{1}{c}{ 5} & \multicolumn{1}{c}{$drop$}  \\ 
  \midrule
\: $\hat{\pi}_u$ &  
   0.440 & 0.109 & 0.198 & 0.205 & 0.048 & 0.000 \\   \midrule
   $\hat{\pi}_{u|1}$  & 0.891 & 0.046 & 0.020 & 0.039 & 0.000 & 0.004 \\ 
     $\hat{\pi}_{u|2}$  & 0.000 & 0.930 & 0.000 & 0.000 & 0.065 & 0.005 \\ 
   $\hat{\pi}_{u|3}$  & 0.002 & 0.009 & 0.871 & 0.012 & 0.083 & 0.023 \\ 
 $\hat{\pi}_{u|4}$& 0.000 & 0.003 & 0.000 & 0.994 & 0.002 & 0.001 \\ 
   $\hat{\pi}_{u|5}$ & 0.000 & 0.016 & 0.013 & 0.060 & 0.639 & 0.272 \\ 
  $\hat{\pi}_{u|drop}$ & 0.000 & 0.000 & 0.000 & 0.000 & 0.000 & 1.000 \\ 
\midrule
 &  \multicolumn{1}{c}{1} & \multicolumn{1}{c}{2} & \multicolumn{1}{c}{3} & \multicolumn{1}{c}{4} &\multicolumn{1}{c}{ 5} & \multicolumn{1}{c}{$drop$}  \\ 
\midrule
\: $\hat{\pi}_u$ &   0.426 & 0.209 & 0.281 & 0.086 & 0.000 & 0.000 \\  
  \midrule
  $\hat{\pi}_{u|1}$ & 0.691 & 0.132 & 0.021 & 0.053 & 0.103 & 0.000 \\ 
  $\hat{\pi}_{u|2}$ & 0.000 & 0.862 & 0.038 & 0.000 & 0.070 & 0.030 \\ 
  $\hat{\pi}_{u|3}$ & 0.000 & 0.000 & 0.806 & 0.000 & 0.183 & 0.011 \\ 
  $\hat{\pi}_{u|4}$  & 0.180 & 0.000 & 0.000 & 0.524 & 0.296 & 0.000 \\ 
  $\hat{\pi}_{u|5}$ & 0.000 & 0.000 & 0.022 & 0.000 & 0.768 & 0.210 \\ 
  $\hat{\pi}_{u|drop}$ & 0.000 & 0.000 & 0.000 & 0.000 & 0.000 & 1.000 \\ 
  \bottomrule  
\end{tabular}
\end{table}
\begin{table}%[ht]
\centering
\renewcommand{\arraystretch}{0.8}
\caption{\em Average of the  %SP: average of 
estimated  initial  and  transition probabilities  
with respect to age: people aged 50 years or older (upper panel)  and 
aged less than 50 years (bottom panel)} \label{tab:trans_age}
\vspace{-0.1cm}
\begin{tabular}{lrrrrrrrr}
 \toprule
&  \multicolumn{6}{c}{$u$}\\
\cline{2-7}
 &  \multicolumn{1}{c}{1} & \multicolumn{1}{c}{2} & \multicolumn{1}{c}{3} & \multicolumn{1}{c}{4} &\multicolumn{1}{c}{ 5} & \multicolumn{1}{c}{$drop$}  \\ 
  \hline
\: $\hat{\pi}_u$ &  
  0.364 & 0.170 & 0.180 & 0.224 & 0.062 & 0.000 \\ \hline
  $\hat{\pi}_{u|1}$ & 0.832 & 0.073 & 0.017 & 0.052 & 0.023 & 0.003 \\ 
  $\hat{\pi}_{u|2}$ & 0.000 & 0.905 & 0.000 & 0.000 & 0.083 & 0.012 \\ 
  $\hat{\pi}_{u|3}$  & 0.001 & 0.000 & 0.827 & 0.019 & 0.124 & 0.029 \\ 
  $\hat{\pi}_{u|4}$ & 0.024 & 0.002 & 0.000 & 0.902 & 0.071 & 0.001 \\ 
  $\hat{\pi}_{u|5}$ & 0.000 & 0.000 & 0.000 & 0.031 & 0.699 & 0.270 \\ 
  $\hat{\pi}_{u|drop}$ & 0.000 & 0.000 & 0.000 & 0.000 & 0.000 & 1.000 \\
\cline{2-7}
 \renewcommand{\arraystretch}{3}
 &  \multicolumn{1}{c}{1} & \multicolumn{1}{c}{2} & \multicolumn{1}{c}{3} & \multicolumn{1}{c}{4} &\multicolumn{1}{c}{ 5} & \multicolumn{1}{c}{$drop$}  \\ 
  \hline
\: $\hat{\pi}_u$ &  
  0.513 & 0.071 & 0.236 & 0.158 & 0.022 & 0.000  \\ 
  \hline
  $\hat{\pi}_{u|1}$ & 0.905 & 0.039 & 0.023 & 0.028 & 0.001 & 0.004 \\ 
  $\hat{\pi}_{u|2}$ & 0.000 & 0.938 & 0.009 & 0.000 & 0.049 & 0.004 \\ 
  $\hat{\pi}_{u|3}$ & 0.003 & 0.015 & 0.903 & 0.001 & 0.064 & 0.014 \\ 
  $\hat{\pi}_{u|4}$ & 0.017 & 0.004 & 0.000 & 0.978 & 0.000 & 0.001 \\ 
   $\hat{\pi}_{u|5}$ & 0.000 & 0.029 & 0.027 & 0.075 & 0.609 & 0.260 \\ 
   $\hat{\pi}_{u|drop}$ & 0.000 & 0.000 & 0.000 & 0.000 & 0.000 & 1.000 \\
  \bottomrule  
\end{tabular}
\end{table}
%
%These probabilities %FB: penso non si capisca These probabilities a cosa si riferisce
%FP cambiato
The average  initial and transition probabilities are reported in Tables \ref{tab:trans_drug}, \ref{tab:trans_gen},  and \ref{tab:trans_age}.
We observe that treated and untreated patients  have an almost equal estimated  probability of dropping out.
For males,  dropout probability is much higher than females when they are in the 2nd state. Moreover, we observe that males have a lower persistence probability in the 1st state with respect to females since  around 13\% of males are estimated to move towards the 2nd state and around 10\% towards the 5th state.
Older patients are less persistent in the first four states than younger patients. 

\begin{figure}[htb]
  \centering
 \caption{{\em Decoded states of  the  HMM with $k=5$ hidden states and dropout state (6th pink)}} 
 \vspace{-0.5cm}
     \includegraphics[width=12cm]{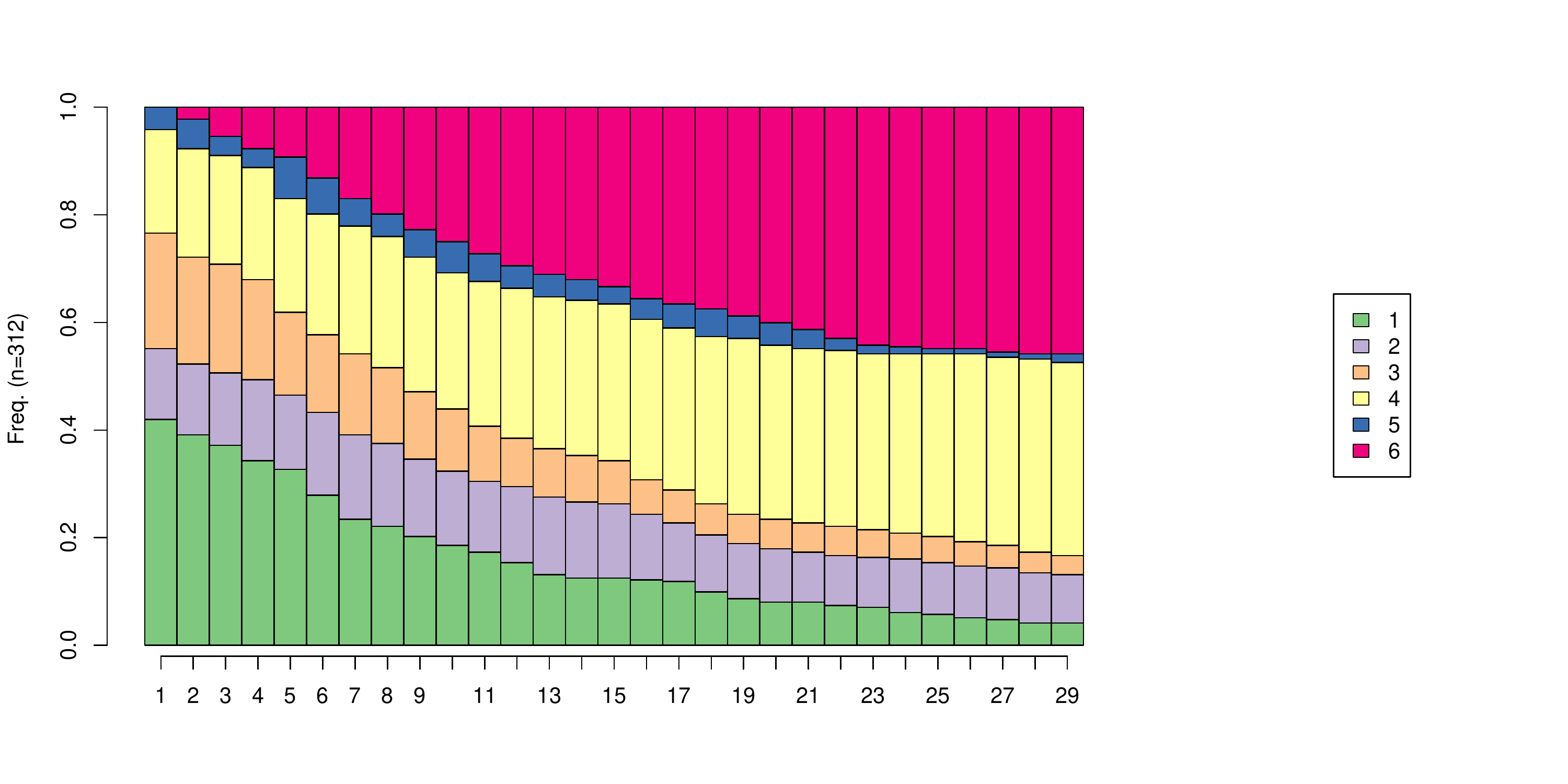}\label{fig:post}
    \end{figure}

\begin{figure}
    \centering
 \subfloat[Not treated]{{\includegraphics[width=8cm]{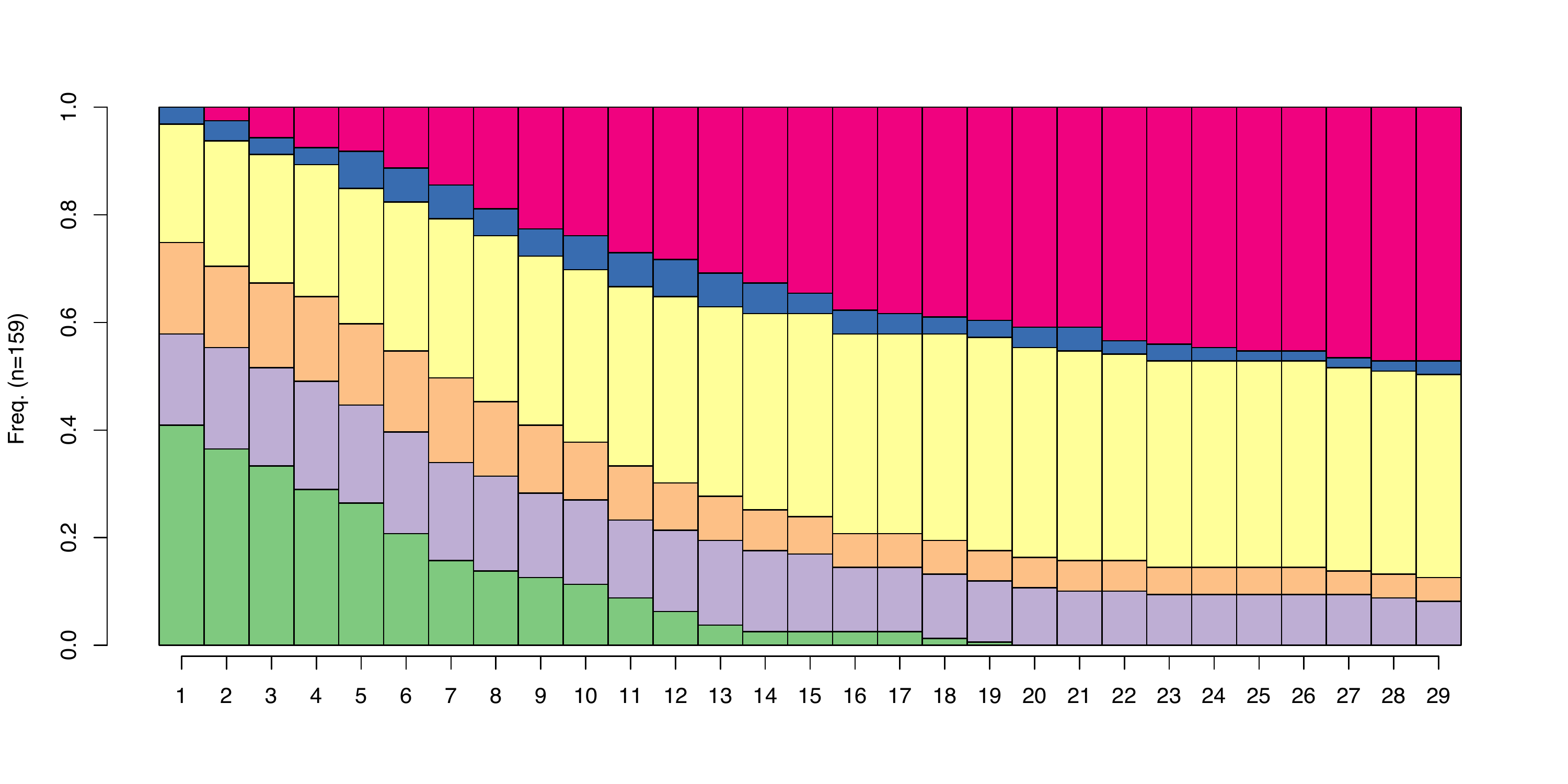} }}
    \subfloat[Treated]{{\includegraphics[width=8cm]{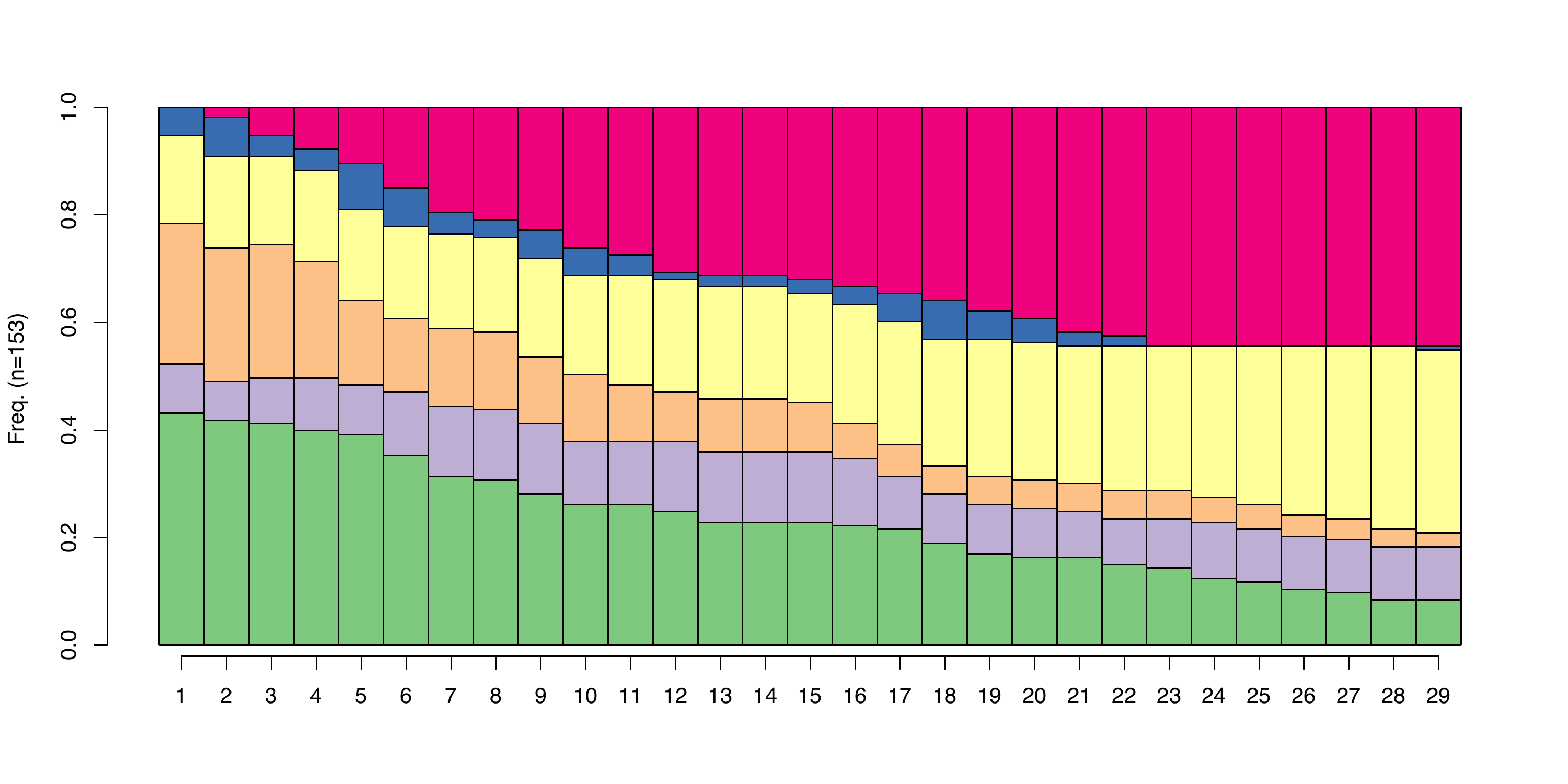}}}
\qquad  
\subfloat[Gender: Female]{{\includegraphics[width=8cm]{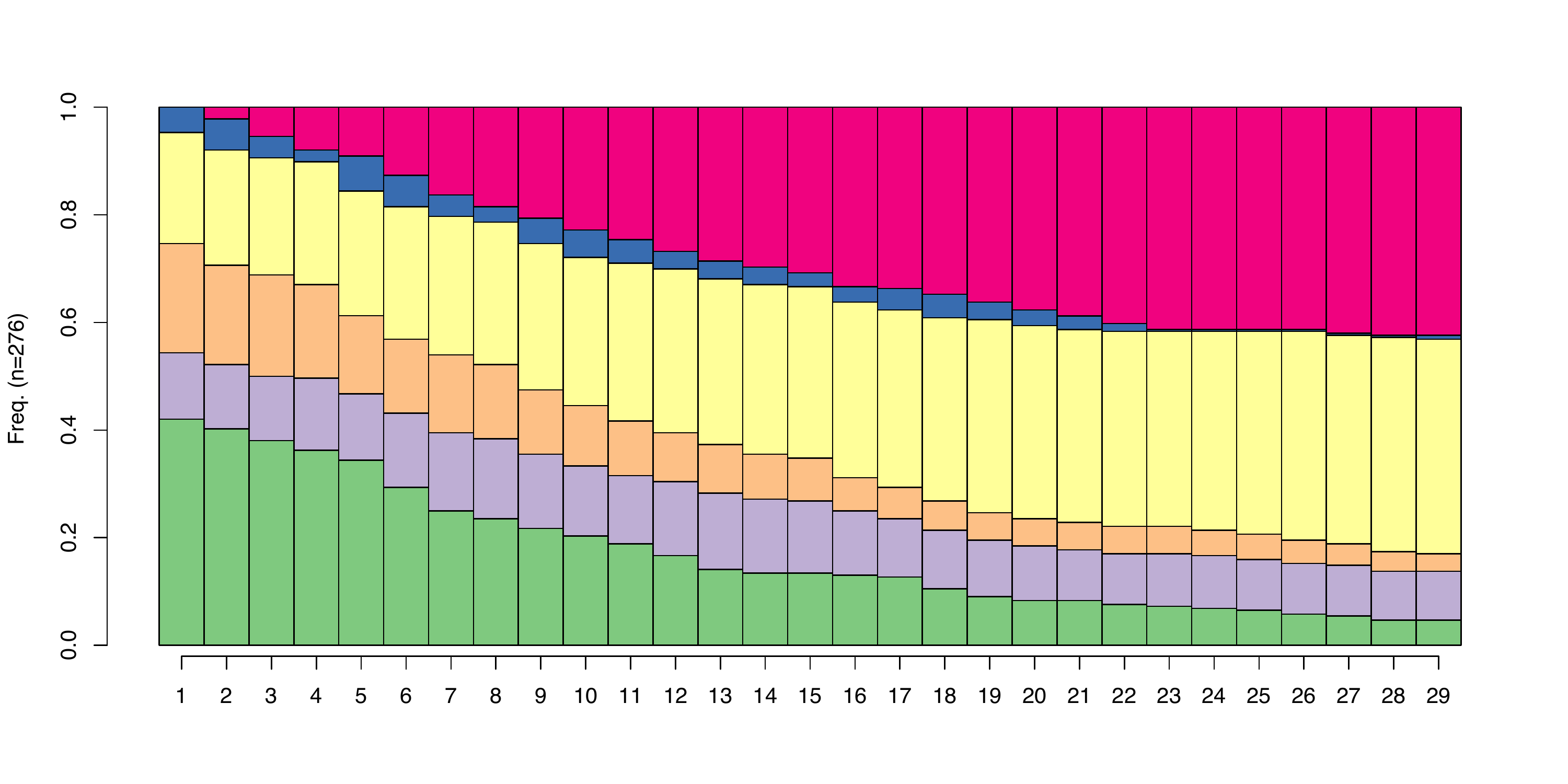} }}
 \subfloat[Gender: Male]{{\includegraphics[width=8cm]{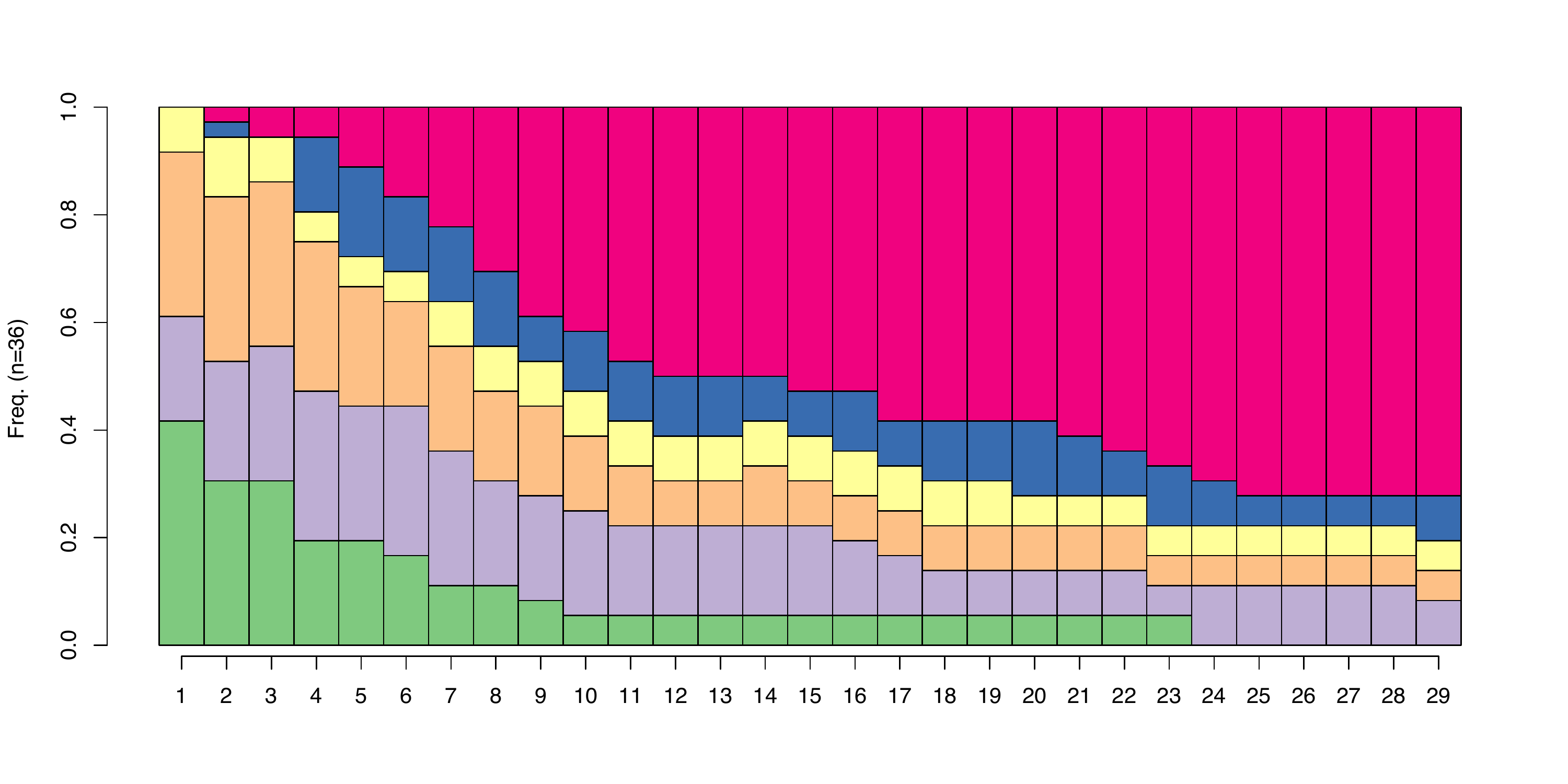} }}
   \qquad  
    \subfloat[Old]{{\includegraphics[width=8cm]{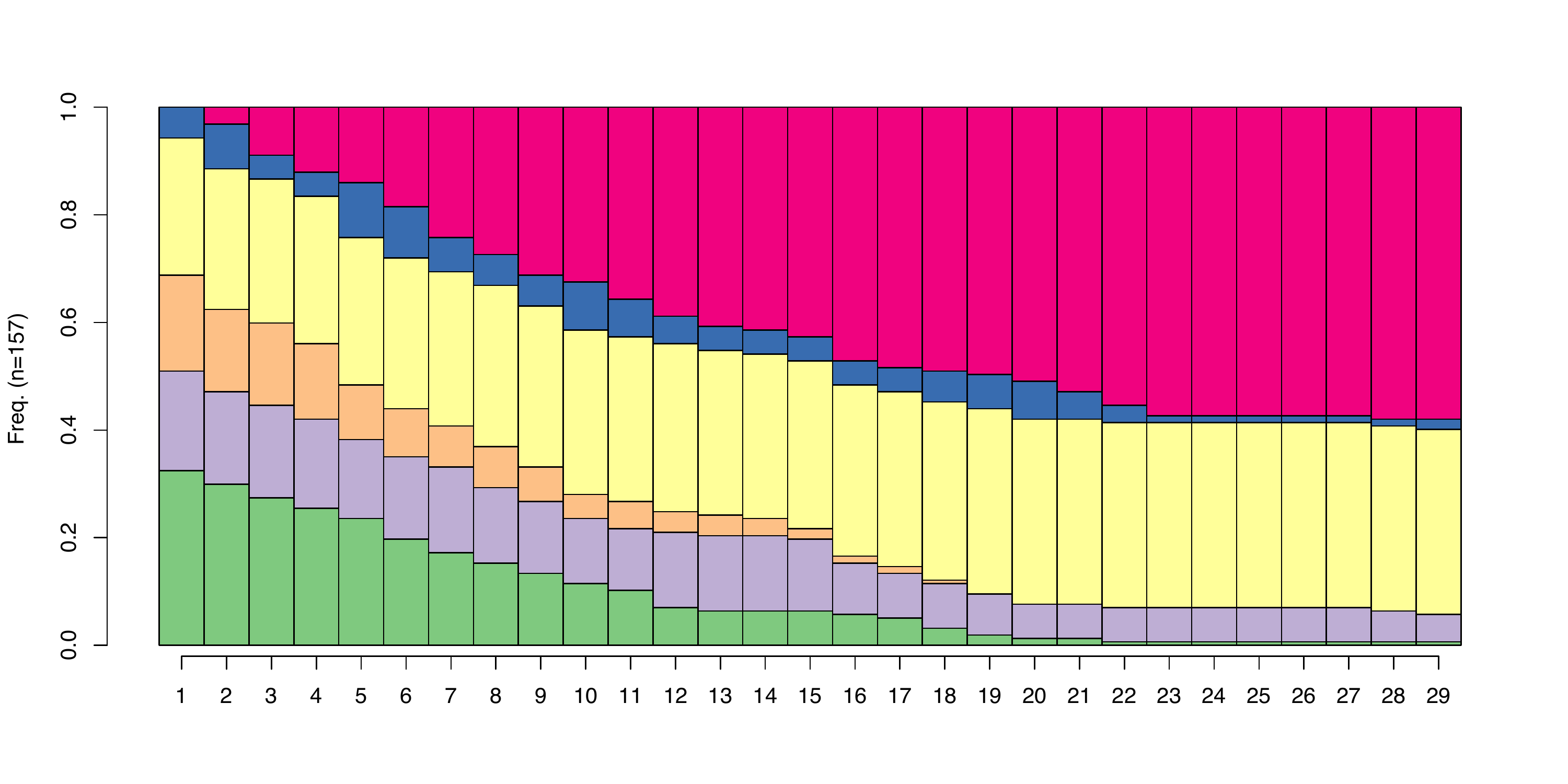} }}
     \subfloat[Young]{{\includegraphics[width=8cm]{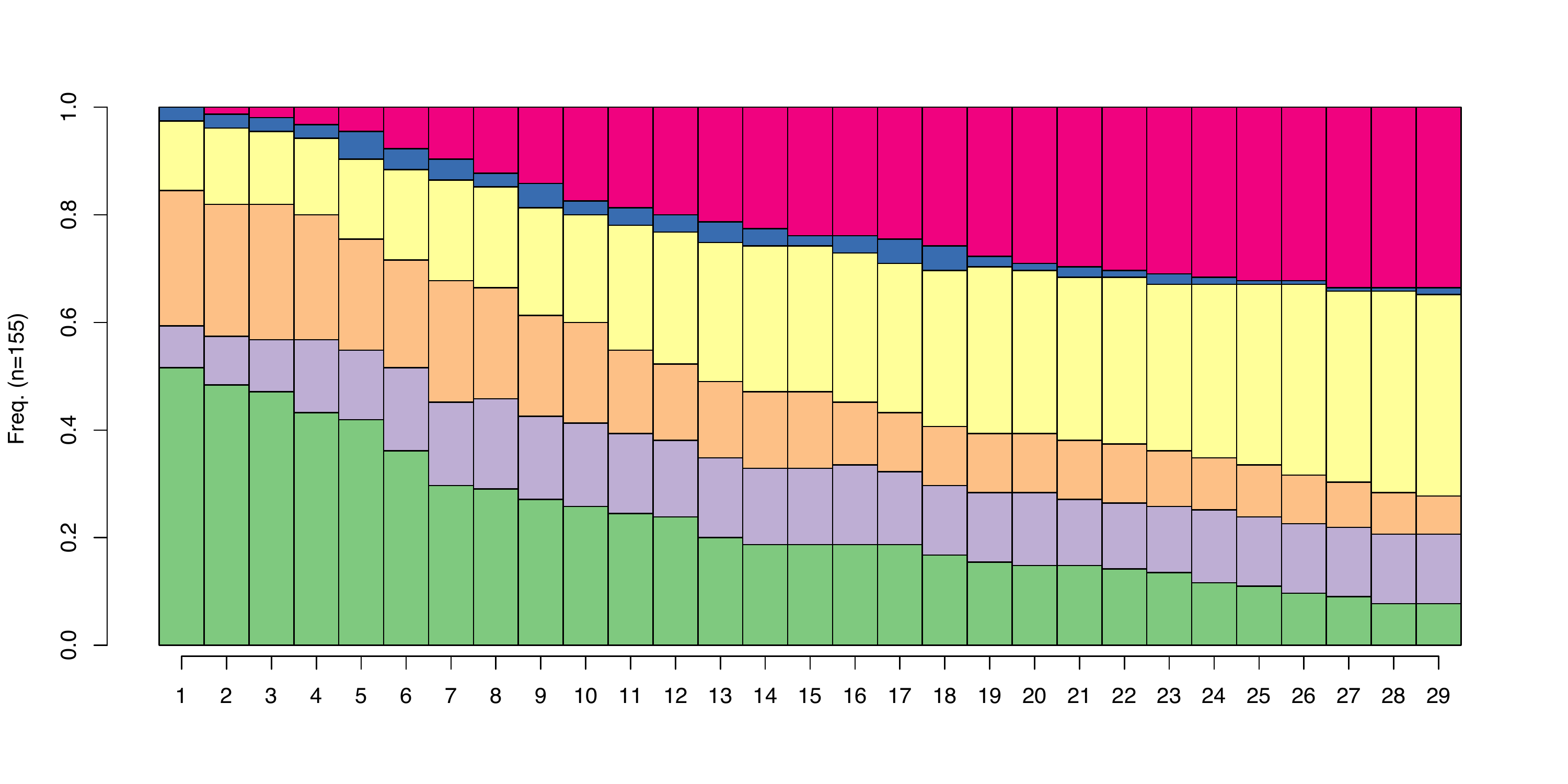} }}
\caption{{\em 
Decoded states of  the proposed HMM with respect to the covariates: drug use, gender, and age; the legend for the states  is the same as that provided in Figure \ref{fig:post}: green 1st, violet 2nd, orange 3rd, yellow 4th, blue 5th, and pink 6th (dropout state)}}
\label{fig:decoc}
\end{figure}

An important aspect of this medical application is  predicting the sequence of latent states to evaluate the time-varying  patient risk of death.
Figure \ref{fig:post} shows the relative frequency of the patients assigned to 
each state at each time occasion  and Figure \ref{fig:decoc} depicts  the proportions of patients allocated in each 
state with respect to the following binary covariates: if they received the D-penicillamine or placebo, females versus males, 
and patients having less than 50 years old at the baseline versus patients older than 50.
Interestingly,  from Figure \ref{fig:post} we notice that it  is possible to predict  the survival curve for all patients by looking at the curve  referred to the dropout state (the pink line).
We recall that this decoding is provided using the estimated posterior probabilities which are directly provided by the proposed EM algorithm  once the  HMM with covariates is estimated. 
From these estimated survival curves, we conclude that 
%FP the non ci vuole il the prima di drug qui
drug use was 
not particularly effective in prolonging survival, males have a higher risk of dying compared to females, especially from the eighth visit, and older patients have a higher risk of dying compared to younger patients from  
 the beginning of the study. 
\section{Conclusions}\label{sec:concl}
In this paper we extend maximum likelihood estimation of a finite mixture model (FMM) of Gaussian distributions with missing at random (MAR) responses to the hidden Markov model (HMM) for continuous multivariate longitudinal data. 
Considering that in this context dropout  typically occurs due to the early termination from the trial, we include  an extra absorbing state in the model so that this type of missing is informative.
Overall, the proposed approach accounts for three types of missing data: the first two are intermittent and correspond to the situation of completely or partially missing responses for a given occasion, while 
the third type of missing data corresponds to dropout.
We implement an exact maximum likelihood inferential approach to deal with the first two types of 
missingness, under the MAR approach.
Estimation is carried out by an extended Expectation-Maximization algorithm implemented  to account for missing values and to include an  extra hidden state for the dropout.
This inferential approach is 
also developed  to estimate the parameters of the model when individual covariates are included  in the distribution of the latent process.
In such a context, it may be of interest to evaluate the effect of these covariates on the transition toward the dropout state. 
We notice 
that in the presence of missing data, the proposed approach also allows us to perform a sort of multiple imputation  so as  to predict the missing responses conditionally or unconditionally to the assigned latent state.

The simulation study allowed %SP:s 
us to conclude that the model parameters are properly estimated even with a relatively large 
proportion of missing responses and dropout.
The application to multivariate data about primary biliary cholangitis referred to several biochemical measurements of liver function  turned out to be particularly challenging.
The data were very sparse due to missing visits of the patients and the fact that some variables were not collected at each visit.
Moreover, 
several dropouts occurred due to death. With the proposed approach, we identified five groups of patients differing for the severity of this rare disease and their transitions across states and towards the dropout  state over time. According to the available covariates, it was possible to predict the distributions of survival times for groups of patients through the decoded states.  
\label{lastpage}
\bibliography{bibliodrop}
\bibliographystyle{apalike}

\clearpage

\end{document}

\label{firstpage}
\maketitle

\section*{Supplementary information}\label{sec:app} 
In the following sections we show some additional details on the historical data used for the applicative example reported in Section 6 of the main paper, and we illustrate supplementary results of the hidden Markov model (HMM) estimated in the paper.  

%FP sezione 1
\section{Data description}\label{ap1}
The applicative example concerns data referred to biochemical measurements related to a double-blinded randomized trial in Primary Biliary Cholangitis (or cirrhosis, PBC) conducted by the Mayo Clinic from 1974 to 1984. 
The data  in different formats are  available  in the library \texttt{JM} \citep{rizo:10} of \texttt{R} \citep{R:21} and on-line  in the  datasets archive at the following website: \url{http://lib.stat.cmu.edu/datasets/pbcseq}.

PBC is a rare but fatal chronic liver disease of unknown cause, with a prevalence of about 50 cases per million inhabitants. The primary pathological event of this disease seems to be the destruction of intralobular bile ducts, mediated by immunological mechanisms.  To be eligible for these trials, patients had to meet well-established clinical, biochemical, serologic, and histologic criteria for PBC.
These data are historical data used in many studies since it has been one of the most extensive existing datasets of a controlled clinical trial on PBC for a long time. 
As remarked in \cite{flem:91}, historical data are important since clinical trials are challenging to complete in rare diseases, and they permit the study of the natural history of the disease before the liver transplant was considered as a  standard practice. Moreover,  disposing of accurate and parsimonious models that can predict survival time based on inexpensive, non-invasive, and readily available measurements is a very relevant feature in clinical science.

The  biochemical variables used as responses in the empirical illustration proposed in Section 
%FP \ref{sec:appl} 
6 of the main paper are the following:
\begin{itemize}
\item   {\em Serum bilirubin}  (Bilir in mg/dl) is a liver bile pigment; normal adult levels range between 0.3-1.0 (min log -1.2, max log 0). Values above 1.2 are synonymous with liver failure;
\item   {\em Serum cholesterol}  (Chol in mg/dl) is a blood lipoprotein; normal adult levels range between 150-199 (min log 4.78, max log 5.39). Borderline values are considered between 200 and 239;
\item   {\em Serum albumin} (Albu in gm/dl) is a protein found in the blood; normal adult levels range between 3.5-5.4 (min log 1.5, max log 1.61). Low albumin values may result from liver malfunction; 
\item   {\em Platelets} (Plat, counts per cubic ml/1000) are components of blood;  normal adult levels range between 150-350 (min log 5.01, max log 6.11); 
\item   {\em Prothrombin} (Prot) is a blood coagulation agent time and time is recorded until a blood sample begins coagulation in a certain laboratory test in seconds; normal adult levels range between 10-13 seconds (min log 2.17, max log 2.45); 
\item   {\em Alkaline phosphatase} (Alka, in U/liter)  is an enzyme; normal adult levels range between 36-150 (min log 3, max log 4.94). High values of alkaline phosphatase can occur in the presence of liver disease; 
\item   {\em Transaminase} (Tran or SGOT in U/ml)   adult normal levels range between 0-35 (min log 2.08, max log 3.69). In liver damage, an increase of transaminase in blood concentration is observed up to five to tenfold in the presence of severe damage. 
\end{itemize}
In particular,   {\em serum bilirubin} level is considered a strong indicator of disease progression and a strong predictor of survival  \citep{flem:91,taa:20}.    {\em Bilirubin} and   {\em albumin} are two of the primary indicators to help evaluate and track the absence of liver diseases. A significantly higher level than bilirubin's standards excreted in bile usually  indicates certain diseases.   {\em Serum albumin} may be harmful to humans having too high or too low circulating levels. Typically, there exist some associations between the levels of    {\em serum bilirubin} and   {\em serum albumin}, and thus it is important to account for a  joint analysis of the longitudinally collected biochemical markers for the  diagnosing of  liver diseases as that proposed in the paper through the multivariate HMM.

\begin{figure}[htb]
  \centering
     \includegraphics[width=13cm]{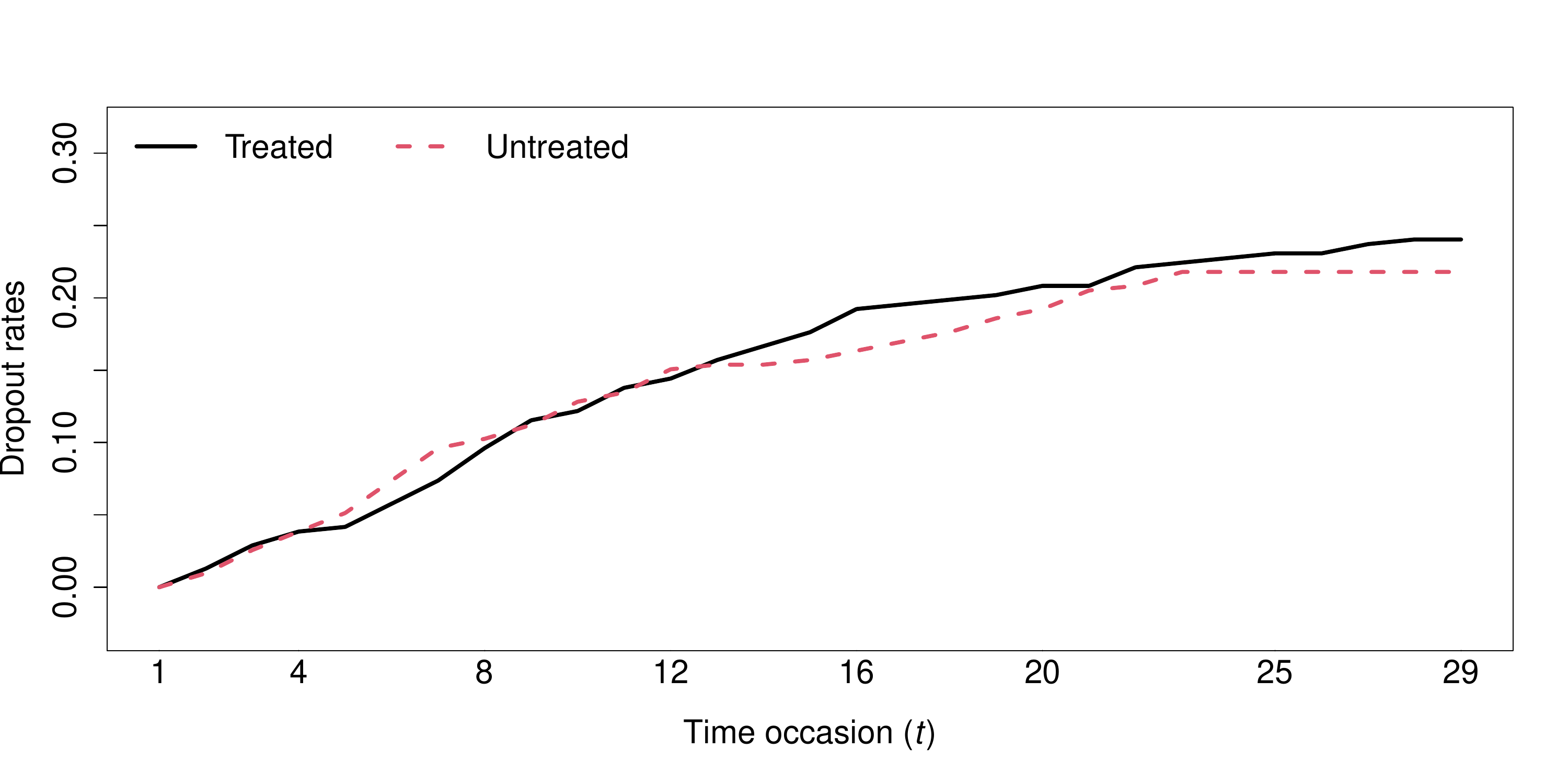}
    \caption{{\em Dropout rates for treated and untreated patients}\label{fig:drop1}}
    \end{figure}

%FP precisato sotto riportando pezzo dal main article
The original clinical protocol for the patients specified visits at 6 months, 1 year, and annually after that. However,  the actual follow-up times varied considerably around the scheduled visits. Therefore, in the analyses we considered  time occasions at 6 months from the baseline, thus accounting for missing observations, missing visits, 
and dropout in a period of 29  time occasions. 
Despite the immunosuppressive properties of the  D-penicillamine, which was the main treatment at that time, no detectable differences were observed between survival time distributions for placebo and treated groups. At the end of the study, 140 patients died, 29 had liver transplantation,  and 143 were still alive.  Figure \ref{fig:drop1} shows the dropout rates for treated and not treated patients over time.

Descriptive statistics of the responses 
for each time occasion are provided in Table \ref{tabAP:desc}.
Due to some atypical observations or outliers in the data as also recently proposed in  \citep{taa:20}   we  considered the natural logarithm of the markers. 
\begin{table}[htb]
\renewcommand{\arraystretch}{.8}
\centering
\tiny{
\caption{{\em Average and standard deviation
%FP (sd)
(sd) of the  responses (in logarithm):  bilirubin, cholesterol,  albumin,  platelets,  prothrombin,  alkaline, and  transaminase from the baseline to the end of the study}}
\vspace{-0.1cm}
{\small
\begin{tabular}{c|rr|rr|rr|rr|rr|rr|rr}
  \hline
  \multicolumn{1}{c}{$t$} &  \multicolumn{1}{c}{Bilir} &  \multicolumn{1}{c}{sd }   &  \multicolumn{1}{c}{Chol }  &  \multicolumn{1}{c}{sd} &
   \multicolumn{1}{c}{Albu} &  \multicolumn{1}{c}{sd} &  \multicolumn{1}{c}{Plat} &  \multicolumn{1}{c}{sd} &  \multicolumn{1}{c}{Prot} &  \multicolumn{1}{c}{sd} 
  &  \multicolumn{1}{c}{Alka} &  \multicolumn{1}{c}{sd} &  \multicolumn{1}{c}{Tran} &  \multicolumn{1}{c}{sd} \\
\hline
1 & 0.57 & 1.03 & 5.80 & 0.43 & 1.25 & 0.13 & 5.50 & 0.40 & 2.37 & 0.08 & 7.27 & 0.72 & 4.71 & 0.45 \\ 
  2 & 0.47 & 1.07 & 6.05 & 0.27 & 1.25 & 0.15 & 5.42 & 0.43 & 2.38 & 0.12 & 7.10 & 0.64 & 4.72 & 0.57 \\ 
  3 & 0.53 & 1.04 & 5.76 & 0.36 & 1.24 & 0.15 & 5.42 & 0.42 & 2.36 & 0.07 & 7.10 & 0.64 & 4.73 & 0.54 \\ 
  4 & 0.48 & 1.01 & 5.78 & 0.34 & 1.16 & 0.21 & 5.27 & 0.43 & 2.39 & 0.07 & 6.73 & 0.75 & 4.55 & 0.58 \\ 
  5 & 0.70 & 1.14 & 5.69 & 0.36 & 1.22 & 0.15 & 5.40 & 0.48 & 2.38 & 0.13 & 7.07 & 0.67 & 4.72 & 0.56 \\ 
  6 & 0.93 & 1.47 & 5.56 & 0.35 & 1.36 & 1.20 & 5.34 & 0.36 & 2.43 & 0.16 & 6.96 & 0.84 & 4.62 & 0.58 \\ 
  7 & 0.59 & 1.11 & 5.72 & 0.36 & 1.23 & 0.13 & 5.39 & 0.48 & 2.37 & 0.08 & 6.97 & 0.61 & 4.65 & 0.56 \\ 
  8 & 0.66 & 1.25 & 5.72 & 0.40 & 1.16 & 0.15 & 5.31 & 0.48 & 2.42 & 0.12 & 6.91 & 0.66 & 4.62 & 0.52 \\ 
  9 & 0.63 & 1.16 & 5.67 & 0.48 & 1.19 & 0.17 & 5.32 & 0.45 & 2.39 & 0.12 & 6.92 & 0.61 & 4.63 & 0.56 \\ 
  10 & 0.40 & 0.92 & 5.50 & 0.29 & 1.23 & 0.15 & 5.25 & 0.47 & 2.38 & 0.10 & 6.93 & 0.62 & 4.59 & 0.43 \\ 
  11 & 0.68 & 1.18 & 5.62 & 0.38 & 1.18 & 0.19 & 5.32 & 0.48 & 2.39 & 0.11 & 6.82 & 0.60 & 4.58 & 0.60 \\ 
  12 & 0.75 & 1.14 & 5.61 & 0.30 & 1.15 & 0.16 & 5.23 & 0.45 & 2.41 & 0.10 & 6.92 & 0.57 & 4.60 & 0.53 \\ 
  13 & 0.73 & 1.17 & 5.61 & 0.31 & 1.17 & 0.15 & 5.27 & 0.47 & 2.40 & 0.09 & 6.73 & 0.58 & 4.59 & 0.65 \\ 
  14 & 0.46 & 1.13 & 5.48 & 0.22 & 1.18 & 0.14 & 5.08 & 0.55 & 2.43 & 0.10 & 6.77 & 0.46 & 4.46 & 0.50 \\ 
  15 & 0.74 & 1.07 & 5.60 & 0.32 & 1.15 & 0.16 & 5.15 & 0.45 & 2.42 & 0.10 & 6.72 & 0.50 & 4.58 & 0.60 \\ 
  16 & 0.54 & 1.15 & 5.54 & 0.35 & 1.13 & 0.26 & 5.23 & 0.49 & 2.40 & 0.07 & 6.77 & 0.52 & 4.46 & 0.55 \\ 
  17 & 0.61 & 1.11 & 5.59 & 0.23 & 1.13 & 0.17 & 5.15 & 0.42 & 2.43 & 0.08 & 6.72 & 0.50 & 4.55 & 0.67 \\ 
  18 & 0.41 & 1.23 & 5.53 & 0.31 & 1.12 & 0.18 & 5.31 & 0.43 & 2.43 & 0.07 & 6.68 & 0.44 & 4.44 & 0.66 \\ 
  19 & 0.62 & 1.15 & 5.58 & 0.32 & 1.12 & 0.20 & 5.17 & 0.46 & 2.48 & 0.21 & 6.66 & 0.59 & 4.51 & 0.71 \\ 
  20 & 1.22 & 1.43 & 5.64 & 0.21 & 0.99 & 0.25 & 5.28 & 0.39 & 2.51 & 0.15 & 6.89 & 0.41 & 4.63 & 0.42 \\ 
  21 & 0.77 & 1.22 & 5.61 & 0.29 & 1.18 & 0.12 & 5.23 & 0.43 & 2.46 & 0.08 & 6.76 & 0.43 & 4.56 & 0.59 \\ 
  22 & 0.21 & 0.37 & 5.78 & 0.06 & 1.26 & 0.04 & 5.14 & 0.50 & 2.38 & 0.02 & 6.82 & 0.25 & 4.32 & 0.25 \\ 
  23 & 0.57 & 1.29 & 5.46 & 0.30 & 1.11 & 0.19 & 5.22 & 0.48 & 2.46 & 0.05 & 6.73 & 0.45 & 4.59 & 0.63 \\ 
  24 & 0.51 & 0.82 & 5.60 & 0.17 & 1.20 & 0.17 & 5.01 & 0.41 & 2.44 & 0.11 & 6.59 & 0.26 & 4.40 & 0.37 \\ 
  25 & 0.42 & 1.03 & 5.57 & 0.20 & 1.71 & 1.89 & 5.13 & 0.45 & 2.47 & 0.06 & 6.73 & 0.53 & 4.38 & 0.59 \\ 
  26 & 1.55 & 2.19 & 5.74 & 0.12 & 1.31 & 0.02 & 5.24 & 0.01 & 2.46 & 0.01 & 6.44 & 0.04 & 5.04 & 0.77 \\ 
  27 & 0.44 & 1.25 & 5.38 & 0.16 & 1.20 & 0.12 & 5.10 & 0.46 & 2.47 & 0.05 & 6.60 & 0.45 & 4.30 & 0.54 \\ 
  28 & 1.22 & 1.60 & 5.80 &0.12  & 1.09 & 0.06 & 4.85 & 0.84 & 2.49 & 0.01 & 7.40 & 1.01 & 4.77 & 1.14 \\ 
  29 & 0.98 & 1.41 & 5.51 & 0.19 & 1.12 & 0.24 & 5.12 & 0.30 & 2.46 & 0.07 & 6.97 & 0.71 & 4.61 & 0.68 \\  \hline 
\end{tabular}\label{tabAP:desc}}}
\end{table}

\begin{figure}[htb] %\label{fig:sp}
    \centering
       \subfloat[Bilirubin]{{\includegraphics[width=6.8cm]{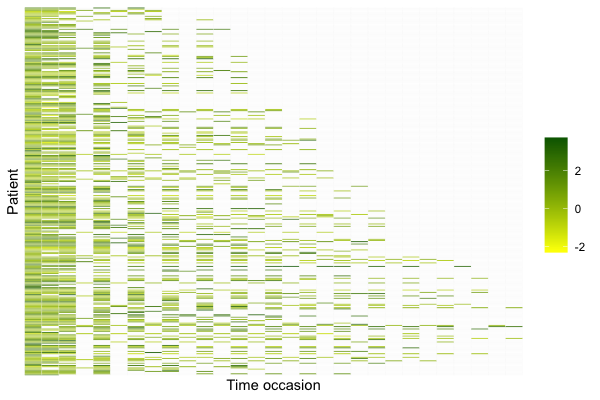} }}%
          \qquad
              \subfloat[Cholesterol]{{\includegraphics[width=6.8cm]{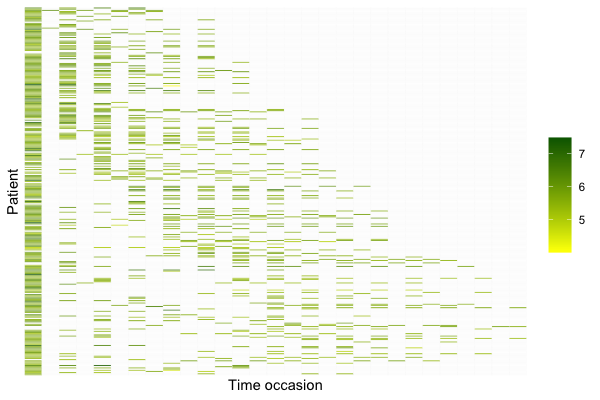} }}%
           \qquad
               \subfloat[Albumin]{{\includegraphics[width=6.8cm]{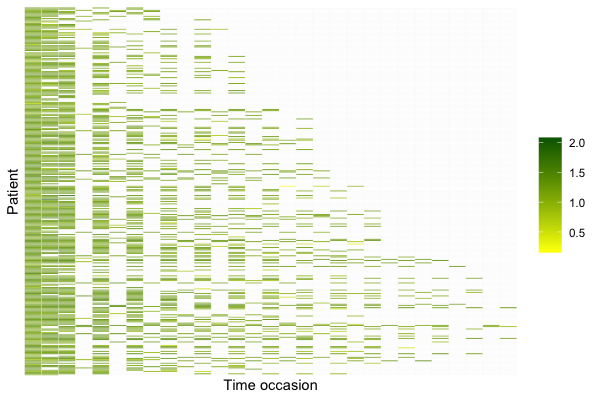} }}
                \qquad
    \subfloat[Platelets]{{\includegraphics[width=6.8cm]{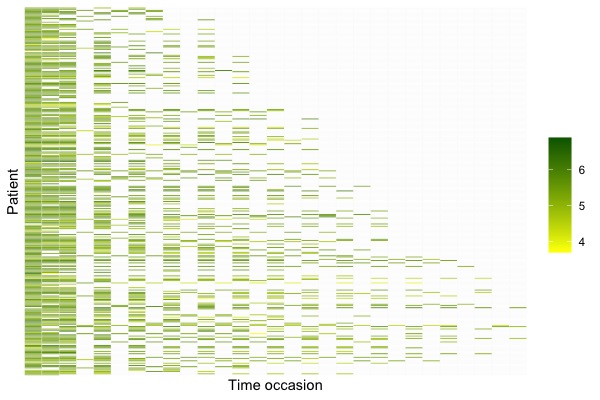} }}%
               \qquad  
    \subfloat[Prothrombin]{{\includegraphics[width=6.8cm]{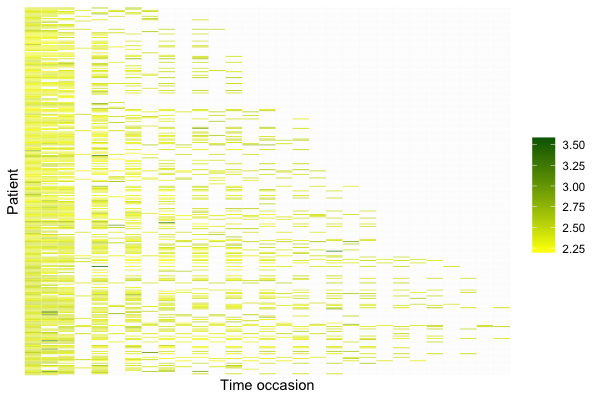} }}
          \qquad  
    \subfloat[Alkaline]{{\includegraphics[width=6.8cm]{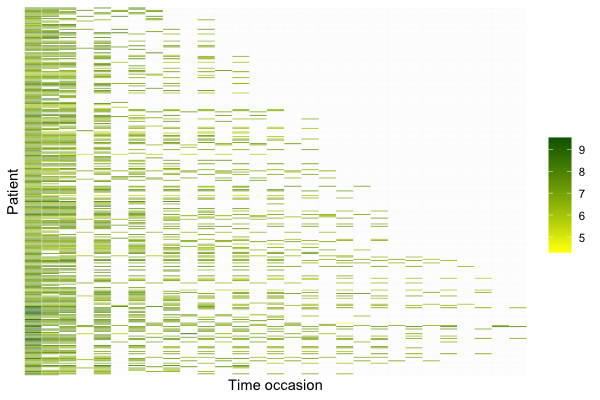} }}
      \qquad  
    \subfloat[Transaminase]{{\includegraphics[width=6.8cm]{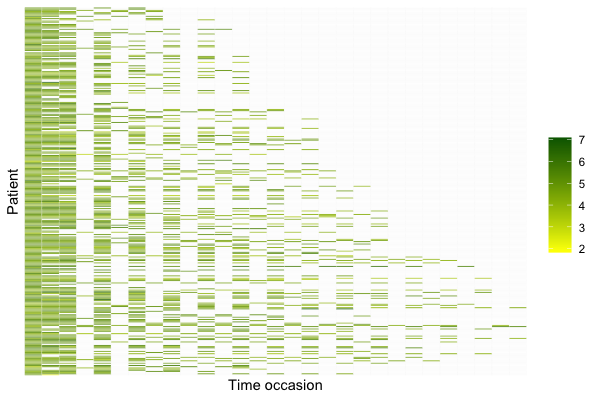} }}
     \qquad
\caption{{\em  Patient's observations for each response variable over the 29 time occasions}\label{fig:las}}%
\end{figure}

Figure  \ref{fig:las}  shows the observed  values for each patient on each response at every time occasion. We notice  that almost all the patients had a record for each variable at the first three visits with the exception of the response referred to {\em cholesterol}. The number of patients with a recorded values after the 11th visit decreased very fast.
Patients made on average 6.2 visits, with an average variation around the mean of   3.8 visits. We observe that {\em bilirubin} 
was recorded on all patients at the baseline except one,  and after the fourth visit it was recorded on only  7\% of patients,   {\em cholesterol} 
was recorded on 90\% of the patients at the baseline and  on only 3\% of the patients on average at the subsequent  visits.

Concerning the available covariates referred to the  drug use,  gender, and age, we notice that females were 88\%, the patient treated  with   D-penicillamine were 51\%. Among  patients with age higher than 50 their  age  ranged  from  50 to 78.5 years with an average age of  58.65 and for patient with an age lower than 50 the range was from 26.5 to 49.5 years with an average age of  41.33.

%FP sezione 2
\section{Additional results}\label{ap2}
In the following we report the estimated standard errors for the 
estimated parameters referred to the effects of the covariates affecting  the initial and transition probabilities 
of the Markov chain 
as in Equations 
(6) and (7) 
of the main paper. 
The following  standard errors  are referred to the estimated coefficients of the HMM with $k=5$ hidden states reported in Tables 10 and 11 of the main paper. These are obtained  (Table  \ref{tab:seboot}) with the non-parametric bootstrap on the basis of  the bootstrap  distribution estimated  with 300 samples. 
Table \ref{tab:seinfo} shows the standard errors  obtained  employing 
the inverse of the expected  information matrix as explained in Section 4.3 of the main paper.  

\begin{table}[htb]
\centering
\renewcommand{\arraystretch}{0.7}
\caption{\em  Standard errors for the estimated effects of the covariates on the initial and transition probabilities of  the  proposed multivariate HMM with $k$ = 5 hidden states obtained with the non-parametric bootstrap }  \label{tab:seboot}
\vspace{-0.3cm}
\begin{tabular}{lrrrrrrr}
\cline{1-6} \cline{1-6}
\vspace*{-0.32cm}\\  & \multicolumn5c{Initial probabilities} 
 & &   \\  \cline{2-6}
Effect &
 \multicolumn1c{$se\hat{\beta}_{12}$} & \multicolumn1c{$ se\hat{\beta}_{13}$} &\multicolumn1c{$se\hat{\beta}_{14}$} & \multicolumn1c{$se\hat{\beta}_{115}$}   &   \\ \cline{1-6}
 Intercept & 1.834 & 1.180 & 2.546 & 3.019 \\ 
  Drug & 0.609 & 0.351 & 0.403 & 1.365 \\ 
  Female & 0.841 & 0.797 & 2.136 & 1.217 \\ 
  Age & 0.027 & 0.019 & 0.020 & 0.048 \\ 
\cline{1-6}	
\vspace*{-0.32cm}\\  & \multicolumn5c{Transition probabilities}  
 & &   \\  \cline{2-6}
\vspace*{-0.32cm}\\ Effect &
 \multicolumn1c{$se\hat{\gamma}_{112}$} & \multicolumn1c{$se\hat{\gamma}_{113}$} &\multicolumn1c{$se \hat{\gamma}_{114}$} & \multicolumn1c{$se\hat{\gamma}_{115}$}   & \multicolumn1c{$se\hat{\gamma}_{11drop}$}  \\ 
\cline{1-6}
Intercept \:\:\:\:\:\ & 1.875\:\:\:\:\:\  & 4.562\:\:\:\:\:\ & 4.253\:\:\:\:\:\ & 5.065 & 2.926 \\ 
Drug\:\:\:\:\ & 0.630\:\:\:\:\:\  & 0.901\:\:\:\:\:\ & 1.751\:\:\:\:\:\ & 3.223  & 3.871 \\ 
  Female\:\:\:\:\:\  & 0.876\:\:\:\:\:\  & 3.455\:\:\:\:\:\  & 3.012\:\:\:\:\:\  & 1.125 & 1.448 \\ 
  Age\:\:\:\:\:\  & 0.030\:\:\:\:\:\  & 0.050\:\:\:\:\:\  & 0.046\:\:\:\:\:\  & 0.157 & 0.083 \\ 
  \cline{1-6}
 Effect &
 \multicolumn1c{$se\hat{\gamma}_{121}$} & \multicolumn1c{$ se\hat{\gamma}_{123}$} &\multicolumn1c{$se\hat{\gamma}_{124}$} & \multicolumn1c{$se\hat{\gamma}_{125}$}  & \multicolumn1c{$se\hat{\gamma}_{12drop}$}   \\ 
\cline{2-6}
Intercept\:\:\:\:\:\  & 1.154\:\:\:\:\:\  & 2.460\:\:\:\:\:\  & 1.891\:\:\:\:\:\  & 1.614 & 8.879 \\ 
  Drug\:\:\:\:\:\  & 0.692\:\:\:\:\:\  & 2.461\:\:\:\:\:\  & 0.481\:\:\:\:\:\  & 0.493 & 3.673 \\ 
  Female\:\:\:\:\:\  & 0.164\:\:\:\:\:\  & 0.941\:\:\:\:\:\  & 0.933\:\:\:\:\:\  & 0.756 & 3.392 \\ 
  Age\:\:\:\:\:\  & 0.048\:\:\:\:\:\  & 0.112\:\:\:\:\:\  & 0.016\:\:\:\:\:\  & 0.024 & 0.125 \\ 
\cline{1-6}
 Effect &
 \multicolumn1c{$se\hat{\gamma}_{131}$} & \multicolumn1c{$se\hat{\gamma}_{132}$} &\multicolumn1c{$se\hat{\gamma}_{134}$} & \multicolumn1c{$se\hat{\gamma}_{135}$}  & \multicolumn1c{$se\hat{\gamma}_{13drop}$}    \\ 
\cline{2-6}	
Intercept\:\:\:\:\:\  & 4.037\:\:\:\:\:\  & 9.702\:\:\:\:\:\  & 14.909\:\:\:\:\:\  & 2.020 & 6.758 \\ 
  Drug\:\:\:\:\:\  & 1.135\:\:\:\:\:\  & 4.920\:\:\:\:\:\  & 4.556\:\:\:\:\:\  & 1.409 & 3.518 \\ 
  Female\:\:\:\:\:\  & 2.435\:\:\:\:\:\  & 6.522\:\:\:\:\:\  & 3.474\:\:\:\:\:\  & 1.357\  & 3.461 \\ 
  Age\:\:\:\:\:\  & 0.108\:\:\:\:\:\  & 0.447\:\:\:\:\:\  & 0.289\:\:\:\:\:\  & 0.025 & 0.117 \\ 
\cline{1-6}
   Effect &
 \multicolumn1c{$se\hat{\gamma}_{141}$} & \multicolumn1c{$se\hat{\gamma}_{142}$} &\multicolumn1c{$se\hat{\gamma}_{143}$} & \multicolumn1c{$se\hat{\gamma}_{145}$}  & \multicolumn1c{$se\hat{\gamma}_{14drop}$}   \\ 
\cline{2-6}
Intercept\:\:\:\:\:\  & 22.283\:\:\:\:\:\  & 6.258\:\:\:\:\:\  & 0.001\:\:\:\:\:\  & 37.205 & 4.918 \\ 
  Drug\:\:\:\:\:\ & 5.533\:\:\:\:\:\  & 3.537\:\:\:\:\:\  & 0.001\:\:\:\:\:\  & 8.747& 2.287 \\ 
  Female\:\:\:\:\:\  & 6.415\:\:\:\:\:\  & 2.504\:\:\:\:\:\  & 0.001\:\:\:\:\:\  & 15.030 & 2.912 \\ 
  Age\:\:\:\:\:\  & 0.358\:\:\:\:\:\  & 0.158\:\:\:\:\:\  & 0.004\:\:\:\:\:\  & 0.607 & 0.110 \\ 
\cline{1-6}
    Effect &
 \multicolumn1c{$se\hat{\gamma}_{151}$} & \multicolumn1c{$se\hat{\gamma}_{152}$} &\multicolumn1c{$se\hat{\gamma}_{153}$} & \multicolumn1c{$se\hat{\gamma}_{154}$}  & \multicolumn1c{$se\hat{\gamma}_{15drop}$}   \\ 
\cline{2-6}
 Intercept\:\:\:\:\:\  & 0.463\:\:\:\:\:\  & 4.046\:\:\:\:\:\  & 13.995\:\:\:\:\:\  & 3.972 & 1.801 \\ 
  Drug\:\:\:\:\:\  & 0.302\:\:\:\:\:\  & 4.705\:\:\:\:\:\  & 9.541\:\:\:\:\:\  & 3.575 & 0.423 \\ 
  Female\:\:\:\:\:\  & 0.467\:\:\:\:\:\  & 4.046\:\:\:\:\:\  & 6.749\:\:\:\:\:\  & 2.393 & 0.725 \\ 
  Age\:\:\:\:\:\  & 0.032\:\:\:\:\:\  & 0.211\:\:\:\:\:\  & 0.494\:\:\:\:\:\  & 0.098 & 0.032 \\  
\cline{1-6}
\end{tabular}\vspace*{5mm}
\end{table}
\begin{table}[htb]
\centering
\renewcommand{\arraystretch}{0.7}
\caption{\em  Standard errors for the estimates of initial and transition probabilities obtained with the information matrix under the  proposed HMM with $k$ = 5 hidden states} \label{tab:seinfo}
\vspace{-0.3cm}
\begin{tabular}{lrrrrrrr}
\cline{1-6} \cline{1-6}
\vspace*{-0.32cm}\\  & \multicolumn5c{Initial probabilities} 
 & &   \\  \cline{2-6}
Effect &
 \multicolumn1c{$se\hat{\beta}_{12}$} & \multicolumn1c{$ se\hat{\beta}_{13}$} &\multicolumn1c{$se\hat{\beta}_{14}$} & \multicolumn1c{$se\hat{\beta}_{115}$}   &   \\ \cline{1-6}
Intercept & 1.438 & 1.008 & 1.273 & 0.909  & \\ 
  Drug & 0.450 & 0.331 & 0.358 & 0.645 & \\ 
  Female & 0.609 & 0.494 & 0.784 & 0.917 &  \\ 
  Age & 0.022 & 0.017 & 0.018 & 0.033 & \\ 
\cline{1-6}	
\vspace*{-0.32cm}\\  & \multicolumn5c{Transition probabilities} 
 & &   \\  \cline{2-6}
\vspace*{-0.32cm}\\ Effect &
 \multicolumn1c{$se\hat{\gamma}_{112}$} & \multicolumn1c{$se\hat{\gamma}_{113}$} &\multicolumn1c{$se \hat{\gamma}_{114}$} & \multicolumn1c{$se\hat{\gamma}_{115}$}   & \multicolumn1c{$se\hat{\gamma}_{11drop}$}  \\ 
\cline{2-6}
Intercept\:\:\:\:\:\ & 1.393\:\:\:\:\:\ & 1.845\:\:\:\:\:\ & 1.694\:\:\:\:\:\ & 0.017 & 1.296 \\ 
  Drug\:\:\:\:\:\ & 0.419\:\:\:\:\:\ & 0.518\:\:\:\:\:\ & 0.574\:\:\:\:\:\ & 0.018 & 1.055 \\ 
  Female\:\:\:\:\:\ & 0.503\:\:\:\:\:\ & 0.952\:\:\:\:\:\ & 0.829\:\:\:\:\:\ & 0.000 & 1.303 \\ 
  Age\:\:\:\:\:\ & 0.024\:\:\:\:\:\ & 0.030\:\:\:\:\:\ & 0.027\:\:\:\:\:\ & 0.021 & 0.052   \\ 	
\cline{1-6}
 Effect &
 \multicolumn1c{$se\hat{\gamma}_{121}$} & \multicolumn1c{$ se\hat{\gamma}_{123}$} &\multicolumn1c{$se\hat{\gamma}_{124}$} & \multicolumn1c{$se\hat{\gamma}_{125}$}  & \multicolumn1c{$se\hat{\gamma}_{12drop}$}   \\ 
\cline{2-6}
Intercept\:\:\:\:\:\ & 0.001\:\:\:\:\:\ & 0.007\:\:\:\:\:\ & 0.001\:\:\:\:\:\ & 1.268 & 0.299 \\ 
  Drug\:\:\:\:\:\ & 0.001\:\:\:\:\:\ & 0.007\:\:\:\:\:\ & 0.001\:\:\:\:\:\ & 0.366 & 1.119 \\ 
  Female\:\:\:\:\:\ & 0.001\:\:\:\:\:\ & 0.001\:\:\:\:\:\ & 0.001\:\:\:\:\:\ & 0.457 & 0.843 \\ 
  Age\:\:\:\:\:\ & 0.001\:\:\:\:\:\ & 0.189\:\:\:\:\:\ & 0.001\:\:\:\:\:\ & 0.019 & 0.017  \\  
\cline{1-6}
 Effect &
 \multicolumn1c{$se\hat{\gamma}_{131}$} & \multicolumn1c{$se\hat{\gamma}_{132}$} &\multicolumn1c{$se\hat{\gamma}_{134}$} & \multicolumn1c{$se\hat{\gamma}_{135}$}  & \multicolumn1c{$se\hat{\gamma}_{13drop}$}    \\ 
\cline{2-6}	
Intercept\:\:\:\:\:\ & 0.045\:\:\:\:\:\ & 0.128\:\:\:\:\:\ & 0.023\:\:\:\:\:\ & 1.417 & 0.853 \\ 
  Drug\:\:\:\:\:\ & 0.001\:\:\:\:\:\ & 1.505\:\:\:\:\:\ & 0.024\:\:\:\:\:\ & 0.520 & 0.910 \\ 
  Female\:\:\:\:\:\ & 0.045\:\:\:\:\:\ & 0.129\:\:\:\:\:\ & 0.026\:\:\:\:\:\ & 0.569 & 1.634 \\ 
  Age\:\:\:\:\:\ & 0.029\:\:\:\:\:\ & 0.031\:\:\:\:\:\ & 0.028\:\:\:\:\:\ & 0.023 & 0.047   \\ 
\cline{1-6}
   Effect &
 \multicolumn1c{$se\hat{\gamma}_{141}$} & \multicolumn1c{$se\hat{\gamma}_{142}$} &\multicolumn1c{$se\hat{\gamma}_{143}$} & \multicolumn1c{$se\hat{\gamma}_{145}$}  & \multicolumn1c{$se\hat{\gamma}_{14drop}$}   \\ 
\cline{2-6}
Intercept\:\:\:\:\:\ & 0.014\:\:\:\:\:\ & 0.908\:\:\:\:\:\ & 0.001\:\:\:\:\:\ & 0.488 & 1.228 \\ 
  Drug\:\:\:\:\:\ & 0.001\:\:\:\:\:\ & 0.909\:\:\:\:\:\ & 0.001\:\:\:\:\:\ & 1.489 & 1.229 \\ 
  Female\:\:\:\:\:\ & 0.001\:\:\:\:\:\ & 0.909\:\:\:\:\:\ & 0.001\:\:\:\:\:\ & 2.027 & 1.229 \\ 
  Age\:\:\:\:\:\ & 0.027\:\:\:\:\:\ & 0.049\:\:\:\:\:\ & 0.001\:\:\:\:\:\ & 0.021 & 0.060 \\ 
\cline{1-6}
    Effect &
 \multicolumn1c{$se\hat{\gamma}_{151}$} & \multicolumn1c{$se\hat{\gamma}_{152}$} &\multicolumn1c{$se\hat{\gamma}_{153}$} & \multicolumn1c{$se\hat{\gamma}_{154}$}  & \multicolumn1c{$se\hat{\gamma}_{15drop}$}   \\ 
\cline{2-6}
	Intercept\:\:\:\:\:\ & 0.001\:\:\:\:\:\ & 0.013\:\:\:\:\:\ & 0.039\:\:\:\:\:\ & 1.068 & 1.109 \\ 
  Drug\:\:\:\:\:\ & 0.001\:\:\:\:\:\ & 0.022\:\:\:\:\:\ & 0.040\:\:\:\:\:\ & 0.988 & 0.332 \\ 
  Female\:\:\:\:\:\ & 0.001\:\:\:\:\:\ & 0.013\:\:\:\:\:\ & 1.683\:\:\:\:\:\ & 1.073 & 0.451 \\ 
  Age\:\:\:\:\:\ & 0.001\:\:\:\:\:\ & 0.070\:\:\:\:\:\ & 0.032\:\:\:\:\:\ & 0.040 & 0.015 \\ 
\cline{1-6}
\end{tabular}\vspace*{5mm}
\end{table}

\label{lastpage}
\bibliography{bibliodrop}
\bibliographystyle{apalike}